\documentclass[10pt,journal,compsoc,twoside]{IEEEtran}

\usepackage{booktabs} % For formal tables
\usepackage{mathptmx}

\usepackage{graphicx}
\usepackage{amsmath}
\usepackage{algorithmic}
\usepackage{array}
\usepackage{url}
\usepackage{fancyhdr}
\usepackage[normalem]{ulem}
\usepackage{hyperref}
\usepackage{enumitem}
\usepackage[english]{babel}
\newif\ifremark
\long\def\remark#1{
\ifremark%
        \begingroup%
        \dimen0=\columnwidth
        \advance\dimen0 by -1in%
        \setbox0=\hbox{\parbox[b]{\dimen0}{\protect\em #1}}
        \dimen1=\ht0\advance\dimen1 by 2pt%
        \dimen2=\dp0\advance\dimen2 by 2pt%
        \vskip 0.25pt%
        \hbox to \columnwidth{%
                \vrule height\dimen1 width 3pt depth\dimen2%
                \hss\copy0\hss%
                \vrule height\dimen1 width 3pt depth\dimen2%
        }%
        \endgroup%
\fi}

\remarktrue

\begin{document}

\title{Accurate and Scalable Many-Node Simulation}

\author{Stijn Eyerman, Wim Heirman, Kristof Du Bois, Ibrahim Hur\\Intel Corporation}

\maketitle
% no keywords
\begin{abstract}
Accurate performance estimation of future many-node machines is challenging because it requires
detailed simulation models of both node and network. However, simulating the full system in detail is unfeasible
in terms of compute and memory resources. State-of-the-art techniques use a two-phase approach
that combines detailed simulation of a single node with network-only simulation of the full system.
We show that these techniques, where the detailed node simulation is done in isolation,
are inaccurate because they ignore two important node-level effects:
compute time variability, and inter-node communication.

We propose a novel three-stage simulation method to allow scalable \emph{and} accurate many-node
simulation, combining native profiling, detailed node simulation and high-level network simulation.
By including timing variability and the impact of external nodes, our method leads to more accurate estimates.
We validate our technique against measurements on a multi-node cluster, and report an average 6.7\%
error on 64 nodes (maximum error of 12\%), compared to on average 27\% error and up to 54\% when
timing variability and the scaling overhead are ignored. At higher node counts,
the prediction error of ignoring variable timings and scaling overhead continues to increase compared to our technique, and may lead to selecting the wrong optimal cluster configuration.

Using our technique, we are able to accurately project performance to thousands
of nodes within a day of simulation time, using only a single or a few simulation hosts.
Our method can be used to quickly explore large many-node design spaces, including node micro-architecture, node count and network configuration.

\end{abstract}

\section{Introduction}

Many-node architectures, such as large HPC systems, cov\-er an important share of the processor market. Being
able to simulate many-node applications, e.g., MPI, is crucial for obtaining accurate projections of
the performance of future architectures. However, no scalable and accurate many-node simulation methodologies
are currently available. Most simulators either target a single node or a few nodes (e.g., gem5~\cite{gem5},
Sniper~\cite{sniper}, MARSS~\cite{marss}), or they perform high-level many-node network simulations with
simple core models (e.g., SST~\cite{sst}, BigSim~\cite{bigsim}). Single-node performance is however impacted
by network transfer latencies and remote memory accesses. Similarly, network performance depends on the timing of
communication events, which is determined by the performance of the nodes. Therefore, isolated and separate
simulations of a single node and the interconnection network cannot provide accurate results.

A straightforward many-node simulation technique would be to simulate all nodes and the interconnection
network in detail~\cite{leon,distgem5}. Apart from the technical issues, this approach is far from scalable: detailed
node simulation is five to six orders of magnitude slower than native execution, and simulating each core in each node requires a huge amount of compute and memory resources. Furthermore, detailed simulation is
sensitive to causality effects, complicating efficient parallel simulation to improve its speed.

In this paper, we study the requirements and propose a novel technique for accurate and scalable simulation of many-node applications.
Our method combines the speed and scalability of native profiling and fast network simulation with the accuracy
of detailed node simulation. We find that in order to obtain accurate estimates, it is crucial to faithfully model timing variability introduced by the microarchitecture, include microarchitectural effects of network communication, and capture the overhead introduced by scaling the problem size.

Using our technique, we are able to predict the performance of five MPI applications executing
on a 64-node (2K cores) cluster to within~6.7\% of real hardware, using just a single simulation host for one
day, compared to an error of 27\% without our technique, and over~60\% when modeling larger systems. In addition, this technique can scale to more than 1,000~nodes and still provide accurate performance projections within a day.

This paper makes the following contributions:
\begin{itemize}[nolistsep]
\item We propose a novel three-phase technique for many-node simulation, targeting both accuracy and
scalability, by combining native profiling, detailed simulation, and high-level network simulation.
\item We accurately model load imbalance and timing variation through capturing distributions and patterns of
compute times during detailed simulation.
\item We augment a detailed node simulator to model the impact on performance of communication and external memory
operations of processes running on other nodes (called \emph{external ranks} in this paper).
\item We use our technique to model the many-node scaling behavior of applications, and measure the effect of
different network bandwidth and topology settings.
\end{itemize}

Before describing our technique in detail, we discuss related work. We then validate our method by comparing it
against measured timings for five relevant MPI applications. We showcase its usefulness by extrapolating our
estimations to 4,000 nodes, and by evaluating the impact of the network parameters.

\section{Related Work}

Our method builds upon both a detailed single-node processor simulator and a network simulator.
We discuss the state of the art for both components in the next sections, and continue with an overview
of prior work on many-node performance projections and the impact of timing variability on multi-node application performance.

\subsection{Single-node simulation}

There are many processor simulators, going from extreme\-ly slow cycle-accurate RTL-level simulators (usually only available to industry),
over detailed academic simulators (e.g., gem5~\cite{gem5}, MARSS~\cite{marss}), to faster approximate
simulators (e.g., Sniper~\cite{sniper}). A lot of research is done on speeding up processor simulation,
such as sampling~\cite{sample_multi,simpoint,smarts}, statistical simulation~\cite{statsim} and using analytical models~\cite{interval_model}.
However, even with these techniques, simulating each node of a many-node application quickly becomes unfeasible.

\subsection{Multi-node simulation}

The growth in data center, cloud and network research has resulted in the development of many simulators that
target many-node simulation. Because of the large difference in application domains, each simulator has specific characteristics
that define its usage model and usability.

Leon et al.~\cite{leon} propose to attach a detailed processor simulator to each simulated node.
A similar approach is taken in dist-gem5~\cite{distgem5}.
While this approach can yield very accurate results, it is in practice limited to a few dozen nodes;
therefore, simulating all nodes in detail becomes unfeasible to model today's HPC machines which can contain 1000s of nodes.
Other simulators provide an infrastructure to simulate nodes and network, and use
flexible node models that can be selected by the user. Node performance can be obtained through native runs~\cite{dimemas,smpi},
analytical core models such as the roofline~\cite{sst} or LogGOPS models~\cite{loggopsim}, or user-provided
timings between communication events~\cite{cloudsim,simgrid,bigsim}.

The application can be modeled as a trace~\cite{dimemas,musa,bigsim} or as an abstract skeleton~\cite{cloudsim,simgrid,sst}.
SMPI~\cite{smpi,smpi2} has a unique approach: it natively executes an unchanged MPI program while running its simulation model.
Traces are collected during the execution of the program, storing communication and computation events. Skeletons are
executable applications that have the same communication pattern as the original application, but without local memory
allocation and computations. Traces have a lower implementation effort compared to skeletons, but they can get very big
and they cannot be (easily) extrapolated to more nodes than the number of nodes that were used to collect them.
Skeletons require more insight into the application, but once a skeleton is made, the simulation can scale to many nodes,
without requiring large trace files.

The technique proposed in our paper has the following goals that motivate our choice of simulators:
\begin{itemize}[nolistsep]
\item Project performance for future architectures. This means we cannot use native profiling, because the hardware is not available yet. Future node architectures can be so disruptive (e.g., Intel Xeon Phi versus Xeon) that the communication and compute pattern is totally different for the same application.
\item Accurately model the microarchitecture timings, the network timings and their interaction. We therefore need detailed microarchitecture and network simulators.
\item Scalable to multiple thousands of nodes. An integrated approach where detailed node and network simulation execute together, is therefore not suited.
\end{itemize}
With these requirements in mind, we developed a methodology with separate node and network simulation, while adding features to model their interaction, such as modeling the impact of communication on memory and cache behavior in the node simulation.
We select an in-house version of Sniper~\cite{sniper} for our detailed node simulator, because it provides a good balance between speed, parallelism and accuracy.
We pick SST~\cite{sst} to model the network, because of its detailed link and switch models and because of its modularity which allows for adding custom timing models.
Comparing the accuracy of different many-node simulators is orthogonal to our work.

\subsection{Impact of timing variability on multi-node performance}

Timing variability, often called jitter, has been recognized as one of the sources of performance loss in HPC
applications~\cite{jitter1,jitter2,jitter3}. Through inter-process communication and synchronization,
all processes will eventually wait on the slowest node. Therefore, even small-probability
events can quickly result in significant degradation on large machines. The main cause reported
in literature are operating system interrupts triggered by page faults and background processes. Therefore, prior work~\cite{jitter_os1,jitter_os2}
has focused on reducing OS jitter. Another cause for timing variability is dynamic frequency scaling~\cite{dvfs}, which can be
solved by fixing the frequency of the nodes.

In this paper we show that the microarchitecture can also be a source of timing variability. Caches, TLBs, prefetchers and branch
predictors lead to variable performance. Multi-core processors share caches, an on-chip network and off-chip memory bandwidth, leading to more variability
due to contention and coherence effects. These effects are expected to increase in the future as more and more cores are integrated on a chip. Although
this variability has a limited impact on the performance of each individual node, the timing differences between nodes become
magnified as we increase the number of nodes to hundreds or thousands.

Timing variability through microarchitectural effects has been studied in the context of real-time systems and worst-case execution
time studies. Proposals were made to reduce this variability~\cite{predictable_perf,chunduri}, at the expense of significantly reducing the average performance.
Performance reduction is not an option for HPC applications, so it is important to model this variability to make accurate many-node
performance estimates.

\section{Many-Node Simulation Framework}

In this section, we discuss our new method of projecting the performance of many-node applications
on future architectures. Before going into detail on this three-stage process, we first explain the
terminology used throughout this section.

\subsection{Terminology}

A many-node application is a single application that is distributed across many (e.g., a few thousand)
interconnected compute nodes that are not in a coherent memory domain. A \emph{node} consists of one or more CPU~sockets,
with multiple processor cores per socket. All cores on a node share the main memory of that node. Nodes are
interconnected through a \emph{network} using a certain technology (Ethernet, InfiniBand, etc.) with a
certain topology (mesh, torus, dragonfly, etc.). The application consists of many \emph{ranks}, where
each rank is a (possibly multi-threaded) process. Ranks communicate through messages (e.g., MPI messages:
send, receive, broadcast, reduce, etc.). Multiple ranks can execute on a single node, e.g., if the number
of threads per rank is smaller than the number of cores per node.

\subsection{Projection Methodology Overview}

\begin{figure*}[tb]
\centering
\includegraphics[width=.7\textwidth]{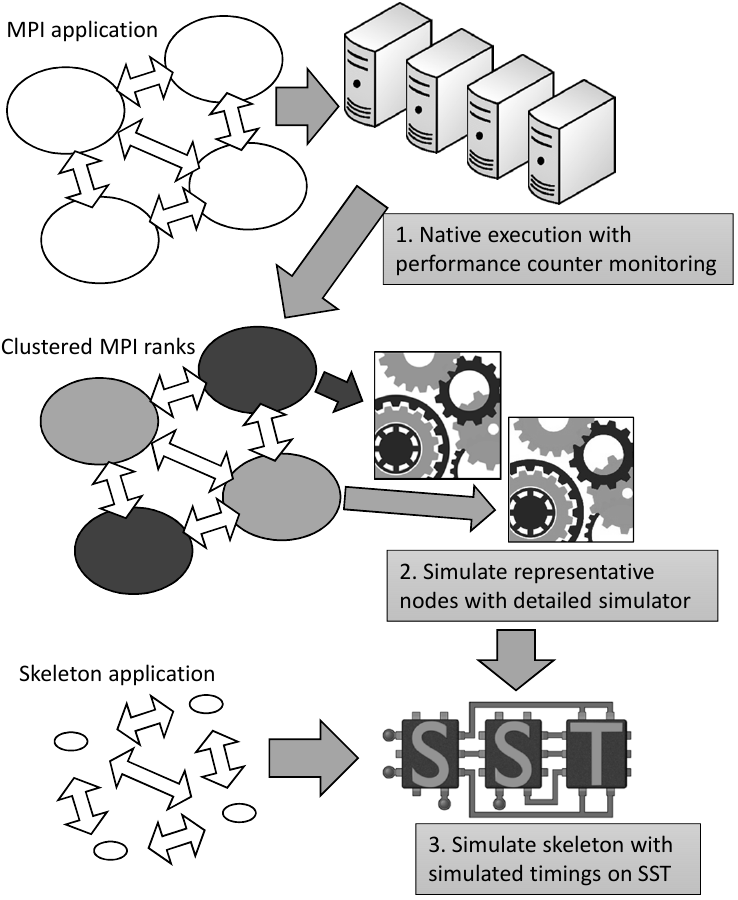}
\caption{The many-node simulation methodology consists of three distinct phases.}
\label{fig:overview}
\end{figure*}

Our proposed many-node simulation method consists of three phases, each gathering information at a different
scale, see Figure~\ref{fig:overview}. The first phase is a native execution of the application on multiple
nodes. The goal of this phase is to collect information on the divergence of rank behavior.

The second phase is a detailed simulation of one or a few nodes executing representative node workloads. 
By only simulating representative nodes, the simulation overhead is greatly reduced. During these
simulations, timings between network events (e.g., MPI calls) are collected. The result of this phase is
a set of distributions of timings between the network events.

In the third phase, we simulate the application on a network simulator for the target number of nodes.
The network simulator does not model the nodes in detail, instead it uses the timings recorded by the
detailed node simulations.
To reduce simulation time and memory pressure, a {\em skeleton} of the application can be simulated instead of the
full application. 
A skeleton is a derivative of the original application, where all local memory allocation
and computation is removed, while retaining all instructions that are required to reconstruct the correct
communication pattern.
The removed memory allocation and computation is replaced by timing calls, generated by the detailed simulations.
Using a skeleton, a many-node simulation can be performed on a single simulation node, reducing the pressure on
compute resources.
However, if compute and memory capacity is abundant, there is no need to construct a skeleton, and the original application can be used.

In the following sections, we go into more detail on each of the three phases.

\subsection{Phase 1: Native Profiling}

Many-node applications are often very regular, i.e., each rank executes the same code, but certain ranks can
behave differently because they are responsible for input and/or output, or they aggregate results from other
ranks.  To detect diverging rank behavior, we execute the many-node application on a real cluster and gather
hardware performance counter information per rank. 
For our setup, we measured instruction count, IPC, number of branches and number of load instructions. 
Although this information depends on the node architecture of the evaluation cluster, the goal of this phase is to detect algorithmic and functional differences between the ranks. 
We assume that these differences are independent of the node architecture, because the partitioning of the application into ranks is determined by the software.
This means that we can use this profile to simulate node architectures that are different from the one we used for profiling, as long as the rank partitioning in the software is not altered.

To get consistent instruction counts, we disable busy waiting in the MPI runtime and threading library, because busy waiting (spinlock)
instructions can artificially increase instruction count.
The number of nodes used in this phase is usually smaller than the final projected many-node computer.
However, the more nodes that can be profiled in this phase, the more potential diverging behavior can be detected.
In our setup, we profiled the applications on 8 nodes, assuming that this number covers most of the diverging rank behavior.
To check this assumption, we redid the profiling for 4 and 16 nodes, which resulted in the exact same conclusion for each application.

After collecting performance counter information, ranks are clustered  using the measured data.
We then select a number of nodes that cover all or most of the clusters.
These nodes are simulated in detail in the next step.
For example, for Caffe (see Section~\ref{sec:setup} for more details on our experimental setup), we find that rank 0 behaves differently from the other ranks.
Rank 0 is the root of the reduction tree, where the new weights for the next neural network training iteration are calculated. 
Even without this semantic information, our performance counter measurements revealed that we should at least simulate a node with rank 0 and a node with another rank.
For the other applications, we find that there are small differences between the ranks, but there is no regular pattern.
Our clustering shows that by simulating the first two nodes, we cover the majority of the different rank behaviors, so we select the first two nodes for the detailed simulation in the next step (for these applications, we run multiple ranks per node because of the limited parallelism with each rank).

The number of clusters determines the balance between accuracy and simulation time: the more clusters, the more
divergent behavior is modeled, but also the more detailed simulations are needed in the next phase.
Note that the communication pattern, and its scaling with increasing node count, is modeled by the communication skeleton in phase 3, so we do not need to profile communication events.
This means we can use existing performance monitoring tools (e.g., the perf tool), eliminating the need for special profiling tools.

If in-depth knowledge of the rank behavior is already available (e.g., because you are the developer of the application), or the application is simple enough to be analyzed by inspecting the code, this phase can be skipped.

\subsection{Phase 2: Detailed Node Simulation}

The next phase consists of a detailed architectural and microarchitectural simulation of a set of representative nodes, as discovered in the previous phase.
It is important to simulate a full node executing as many ranks and threads as one node of the target system, because many resources
on a node are shared (e.g., the network-on-chip, shared caches, memory controllers, etc.).
This ensures that the performance impact and variability due to contention for these resources is modeled accurately.
Additionally, execution-driven simulation is preferred over trace-based approaches (such as~\cite{musa}), to ensure synchronization effects
internal to a node (including synchronization inside multi-threaded ranks and between ranks on the same node) are all taken into account accurately.

The goal of this stage is to capture the timing between network events (e.g., MPI calls). We therefore annotate the
application with {\em markers}, which indicate the places in the code where we want to collect timings,
and which instruct the simulator to write out the simulated timestamp at these points in the code.
Adding these markers can easily be automated by including them in the MPI library.
The timing of the communication must be excluded, as this will be modeled by the network simulator in the next phase.
Therefore, markers are added both before and after each MPI call: the time needed to reach an MPI call is the time between the
end of the previous MPI call until the start of the next MPI call.
We support multithreaded ranks as long as the communication (MPI calls) are done by a single thread per rank, as is the case for most well-designed applications. 

Once the simulation finishes, we collect the timings of the markers. Markers are often inside a loop, meaning that a
marker can be executed multiple times. We group and analyze the timings of multiple executions of the same marker, and cluster timings that have a multimodal distribution (multiple distinct distributions).
For each cluster, we record the cumulative distribution (in maximum 100 bins, to limit the size of the timing profile). 
We also detect timing patterns. For example, for one compute phase of HPCG, the tool detects the following pattern of clusters:
0 1 2 2 2 1 0, repeated 50 times. Timing 0 is larger than timing 1, which is larger than timing 2.
Manual verification of the application source code confirmed that our tool automatically found the correct pattern:
this method was called first with the full matrix, then with a matrix with each dimension halved,
then again halved, and once more halved, after which the matrix was scaled up again 3 times (the timing difference between the two smallest sizes is too small to make it two distinct clusters).
These patterns and distributions are replayed in the skeleton (see next section).

In this phase, we simulate one node in isolation, because our simulator can only simulate one node, as do most architectural simulators. 
However, communication from and to ranks on other nodes might impact
performance, e.g., by causing external memory operations. In Section~\ref{sec:comm}, we discuss additions to the detailed
node simulator to model the impact of external communication on performance.

\subsection{Phase 3: Many-Node Network Simulation}

In this phase, we perform the final many-node simulation. Here, we do not model the nodes in detail,
but we use the collected distributions of timings from the previous phase.
We do perform detailed simulation of the interconnection network between the nodes, including the effects
of link latencies, routing, congestion, buffer overflows, etc.
While avoiding the need for detailed node simulation in this phase speeds up simulation significantly,
retaining all computation and memory
allocation in the application can still create too much overhead, especially if the number of host machines available to run the simulation
is much smaller than the number of simulated nodes. Therefore, we use a skeleton of the application: all
computation and memory allocation on local data is removed, as long as it is not needed to reproduce the correct
communication pattern. Skeletonization is a common approach in network simulation as an alternative to synthetic
traffic or trace replay. We refer to the SST documentation~\cite{sst} for more information.
We implemented the skeletons manually, automatic skeletonization is subject of ongoing research~\cite{skeleton}.
The SST network simulator models the most common MPI calls (including collective and point-to-point communication, as well as non-blocking MPI calls).

Note that we choose to use skeletons instead of traces because they can scale easily to many nodes without requiring large traces, and because they also faithfully model changes in the communication pattern if per-node performance changes while exploring different node micro-architectures.
Nevertheless, the timings generated by the first two phases can also be used in a trace-based network simulator.
Furthermore, if the compute or memory requirements of the application are low, or there are abundant compute and memory resources (e.g., a full cluster is available for simulation), skeletonization might not be needed. Instead, the original application can be used in the simulator.

\subsection{Modeling Timing Variation}\label{sec:variable}

During phase 2 (detailed simulation), we collect the cumulative distribution of the timings of each compute phase. We replay these timings by randomly drawing a timing from this distribution (Monte Carlo simulation). We also make sure we replay potential patterns faithfully, in order to mimic the application behavior as accurately as possible.

\begin{figure}[tb]
\centering
\includegraphics[width=0.5\columnwidth]{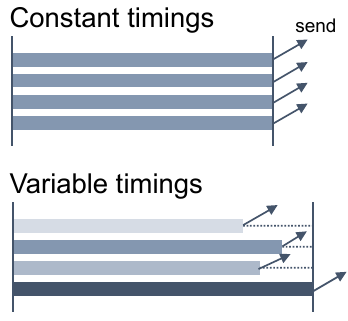}
\caption{Illustration of a collective operation after a compute phase, using constant and variable time modeling. Variable timings more realistically model synchronization delays and network injection patterns, avoiding artificial traffic spikes.}
\label{fig:variable}
\end{figure}

Figure~\ref{fig:variable} shows the importance of modeling variable timings. It represents a computation phase, ending with
a collective operation (e.g., {\tt MPI\_Allreduce}). With constant timings, all compute takes the same amount of time, and all send
operations occur at the same point in time. This causes an artificial traffic spike, leading to (often unrealistic) network
congestion. With variable timings, the total execution time becomes longer, despite the fact that the average timing is the same.
That is because the execution time is determined by the slowest thread. Using constant
timings does not capture this behavior. Furthermore, the send operations are now more spread out over time, leading to a more
realistic network injection behavior.

In our hardware measurements, we make sure to reduce OS noise as much as possible, and we fix the frequencies of all nodes.
Furthermore, our detailed simulator simulates user-level instructions only, so no OS behavior is simulated. All timing
variability stems from microarchitectural effects in caches, TLBs, predictors and the network-on-chip. As we will show in the
evaluation section, these effects lead to significant performance impacts.

\subsection{Modeling External Communication}\label{sec:comm}

In the second phase of our methodology, a single node is simulated to obtain timings between communication events.
However, simulating a single node of a many-node application in isolation can produce inaccurate timings for two reasons:

\begin{enumerate}[nolistsep]

\item \emph{Problem size scaling:} we need to simulate a single node as if it is part of a many-node application,
i.e., we have to scale the problem size to a single node. This can be a problem for some applications, e.g., when
there are no small input sets. Furthermore, ranks might have copies of other ranks' local data that is needed for
their calculations. Scaling the input to a single node might not model memory operations and computation on this
global data.

\item \emph{External communication:} when set up for a single node, no external communication occurs.
External communication is done through buffers in memory, which cause extra memory operations that can have an
impact on cache behavior and memory bandwidth usage. In particular, InfiniBand uses remote direct memory access
(RDMA)~\cite{rdma}, which means that external processes can directly access the memory of a node, e.g.,
the sending node writes a message in the memory of the receiving node.

\end{enumerate}

In order to model the impact of external communication on single-node performance more accurately, we augment
our node-level simulator with two extra features:

\begin{figure}[tb]
\centering
\includegraphics[width=.8\columnwidth]{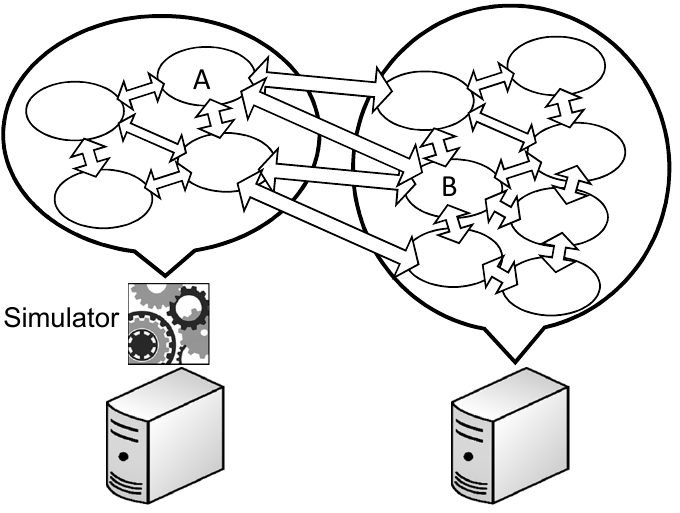}
\caption{Node simulation with external ranks: 4 ranks are simulated in detail, the other ranks run natively on a different host.}
\label{fig:external}
\end{figure}

\begin{enumerate}[nolistsep]

\item \emph{The ability to run external MPI ranks in addition to the ranks on the simulated node.} These
extra ranks run natively, but they do communicate and synchronize with the simulated ranks, see Figure~\ref{fig:external}.
Because the simulator is much slower than native execution, the external ranks wait most of the time until the simulated
ranks have reached a synchronization point. Therefore, these extra ranks do not consume much CPU time, meaning that it is
possible to run many of them on a single or a few node(s). This feature enables running a large-scale many-node application
while only simulating a single node, removing the need to figure out how to scale the problem size. 
Note that during simulation in the second phase, we only record the timings of the compute phases, excluding the communication time.
The progress of the external ranks has no impact on these timings.

\item \emph{Injection of external memory operations.} Whenever a simulated rank (e.g., rank A in Figure~\ref{fig:external}) sends data to an external
rank (e.g., rank B), it writes its send data into a buffer. This buffer is then read by the external rank,
but (because the external rank is invisible to the simulator) memory and cache effects
due to this read operation do not affect simulation state such as cache contents and bandwidth utilization.
We add a mechanism that automatically injects read operations to send buffers, and write operations to receive buffers,
to cause the corresponding memory traffic and coherency updates inside the simulator. This allows for modeling this extra traffic without performing full-system simulation.

\end{enumerate}

Together, these features allow for a more accurate single-node simulation. The first feature (external ranks) can be used
without the second, e.g., if the impact of external communication on memory performance is expected to be negligible.
The second feature (memory operation injection) is only possible when external ranks are included.
\emph{We are not aware of another user-level architectural simulator that has these features.}

\subsection{Discussion and Limitations}

As discussed before, simulating many-node applications is a challenging task, especially if resources in space and time are limited \emph{and} good accuracy is required.
The method presented here represents our current best effort to reach these goals.
Admittedly, it requires some user intervention and it is not applicable to all possible many-node applications.
In this section, we discuss some of the limitations and how they can be overcome.

\paragraph*{Manual user interventions}
The proposed method requires some effort from the user to set up the simulation, such as collecting performance counters from a native run, inserting markers in the detailed simulation phase, and implementing a skeleton.
The first two can be easily automated by using scripts and adding MPI library wrappers.
The largest effort is put into writing the skeleton.
This should preferably be done by the application developer, who knows the communication pattern of the application.
Writing a skeleton is a one-time effort that is justifiable for widely used HPC kernels and application-optimized supercomputer configurations, for which extensive application study is required anyway.
As an example, implementing a skeleton for Caffe deep neural network training (our most complex application) took about one man-week, without initial knowledge of the code and with little experience in writing skeletons.
Taking into account that some HPC kernels already exist for a few decades, and that procurement projects for supercomputers can take multiple months, this cost acceptable.
In fact, many skeletons of common HPC kernels do already exist: we used an existing skeleton for three of the five evaluated applications (HPCG, SNAP and MILC).

\paragraph*{Modeling future architectures}
Our technique is targeted at modeling future architectures, e.g., for exploring the design space of the next generation of processors or network infrastructure.
The ability to model future architectures is ensured by our detailed node simulator (phase 2) and network simulator (phase 3). 
By using flexible and parameterized core and network models, included in our processor simulator Sniper and network simulator SST (see Section~\ref{sec:setup}), many different designs can be evaluated. 
In this paper, we validate our predictions to an existing supercomputer, in order to show its accuracy, something we cannot do for future designs.

A less straightforward phase for modeling future architectures is the first phase, i.e., the native profiling.
Native profiling needs to be done on an existing processor, which is different from the processor we want to model.
We are convinced that if the next generation processor is an evolution of the current processor (e.g., they have a similar instruction set), and if the source code is not changed, then the functional behavior of the ranks will not change much, and the profile information is still valid.
In case the future processor is very disruptive compared to existing processors (e.g., a GPU versus a CPU), and native execution on a similar architecture is not possible, a fast functional simulator can be used.
Functional simulation is slower than native execution, but much faster than detailed architectural simulation.
It also provides most of the counters needed for the clustering: instruction, branch and load count.

We would like to stress that the first phase of our methodology is not a crucial step in obtaining the final performance prediction.
It is needed to reduce the number of simulations in the second phase.
All timing simulations are done in the second and third phase.
The first phase can be replaced by a manual analysis of the application or picking a random sample of nodes.

\paragraph*{Strong scaling}
The simulated compute timings from the second phase can be used to simulate multiple node counts in the third phase, as long as the problem size per node remains constant, i.e., weak scaling.
In case of strong scaling, the detailed simulations in the second phase need to be redone in order to obtain timing information for every per-node problem size.
However, the typical applications for our technique (procurement, network design, etc.) do not consider strong scaling from 1 to thousands of nodes, they rather require performance predictions for one or a few node counts (e.g., should we buy 1,000 powerful nodes or 2,000 cheaper nodes?).
Our technique is therefore not targeted at extensive application scaling studies, but rather at performance predictions for one or a few fixed node count(s) and/or a weak scaling study.

\paragraph*{Non-MPI many-node applications}
Our simulation method is focused at MPI many-node applications.
This choice was impacted by the availability of MPI applications and the ability of SST, our network simulator, to simulate MPI communication primitives.
If a network simulator is available or extended for simulating other many-node paradigms, such as MapReduce~\cite{mapreduce} or gRPC~\cite{grpc}, our method can be also used to model these applications.

In other cases, synchronization between threads and ranks is low or non-existing, e.g., web applications with many independent requests that need multiple services running on different nodes.
The performance of these applications is less sensitive to variations in individual threads, and therefore our technique might have too much overhead for the needed accuracy.
Other simulators might be more relevant for this type of applications.

\section{Experimental Setup}\label{sec:setup}

\begin{table}[!t]
\renewcommand{\arraystretch}{1.1}
\caption{Applications and their problem sizes. }
\label{tbl:apps}
\centering
\begin{tabular}{llll}
\hline
Application & Problem size & OMP threads &MPI ranks\\
&per rank&per rank& per node\\
\hline
HPCG \cite{hpcg} & 64,64,64 & 1 & 16\\
SNAP \cite{SNAP} & 1000,4,4  & 1 & 32\\
MILC \cite{MILC} & 6,6,6,6 &4 & 8 \\
FFT2d \cite{FFTW05} & 32K numbers & 1 & 32\\
Intel Caffe \cite{caffe} & 128 images & 34 & 1\\
\hline
\end{tabular}
\end{table}

We evaluate our technique using four HPC applications and one machine learning application (deep neural network training using Caffe), see Table~\ref{tbl:apps}. This set of applications
covers a range of diverse application behaviors: from compute-intensive (SNAP, Caffe), over memory-intensive
(HPCG, FFT2d) to commu\-ni\-ca\-tion-in\-ten\-sive (MILC).
For all applications, we use a fixed data set size per rank when increasing the number of ranks ({\em weak scaling}).
Consequently, a constant execution time indicates perfect scaling.
Because the OpenMP (OMP) scaling of the HPC applications is fairly weak (or non-existing), we execute multiple ranks on one node.

We execute all applications natively on a cluster of 1 to 64 dual-socket Intel Xeon (Broadwell) nodes. Each node has 36~cores running at
2.3~GHz. 
The nodes are connected using Intel Omnipath (100 Gb/s) in a fat tree topology. 
We use the 8-node hardware runs as a profiling step for phase~1.
The other runs are used to determine the accuracy of our projections.

The applications are simulated on an in-house version of the Sniper
multicore simulator~\cite{sniper} 
for obtaining the timing between communication calls (phase 2). 
We configured Sniper to closely model the Xeon Broadwell nodes of the evaluated cluster. 
To be able to simulate a full node (36 cores)
for at least a few billion instructions, Sniper models some of the core components at a higher level, making
it fast, but not fully cycle-accurate. The average error on the simulated execution time for a single node is 5.3\%
over the evaluated applications. We added the ability of running external ranks and injecting external memory
operations. On average, simulating a single node takes 12~hours and covers between 20~and 250~billion instructions.

The final multi-node simulation (phase 3) is done using the SST simulator~\cite{sst}. SST is a modular simulation
framework that allows for adding custom models. It is shipped with a detailed network simulator, and capabilities
to model the most common MPI operations. The core models are less detailed, but user-defined compute times can be
included, which we use in our methodology. 
In particular, we use the ember element for implementing the skeletons (called `motifs' in ember). The simulated network is a fat tree with 100 Gb/s links.

Simulating a small network ($<$ 100 nodes) using SST takes a few minutes, which
means that for small scales, the evaluation time of our methodology is dominated by the detailed simulations.
However, after obtaining the timings through detailed simulation, we can perform multiple network simulations
with different parameters, amortizing the cost of detailed simulation. Simulating large networks ($>$ 1,000 nodes)
can be done within a few hours, which means that the full methodology (native execution, detailed simulation, and
network simulation) can estimate performance for large systems within one day.

\section{Validation}

We validate the accuracy of our technique by comparing the projected results to the timings measured
on real hardware. Most graphs presented below contain five curves, representing the following configurations:
\begin{itemize}[nolistsep]
\item \emph{Measured:} Total execution time on real hardware.
\item \emph{No external -- constant (no ext-ct):} Estimated execution time using our technique,
but without modeling timing variability and without adding external ranks in the single-node detailed
simulation. This corresponds to the most accurate method reported in literature.
\item \emph{No external -- variable (no ext-var):} Estimated execution time using our technique
including modeling variable timings, but without external ranks.
\item \emph{External -- constant (ext-ct):} Our technique with constant timings, but with external
ranks and external memory operations modeled during single-node simulation.
\item \emph{External -- variable (ext-var):} Our full proposal: modeling timing variability and
adding external ranks.
\end{itemize}

We begin by showing the accuracy of our technique as a function of the number of nodes (from 1 to 64).
Next, we break down the measured and estimated execution time into computation and communication components,
and show that individual components are accurately estimated.

\subsection{Scaling Projection Accuracy}\label{sec:accuracy}

\begin{figure}[ptb]
\centering
\includegraphics[width=\columnwidth]{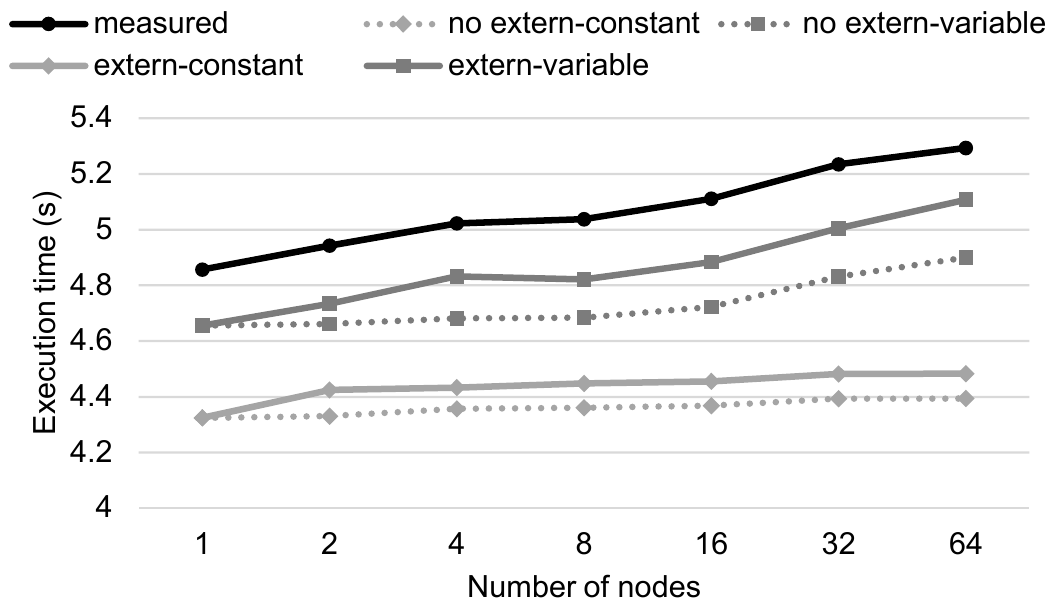}\\
(a) HPCG\\ 
\vspace{4mm}
\includegraphics[width=\columnwidth]{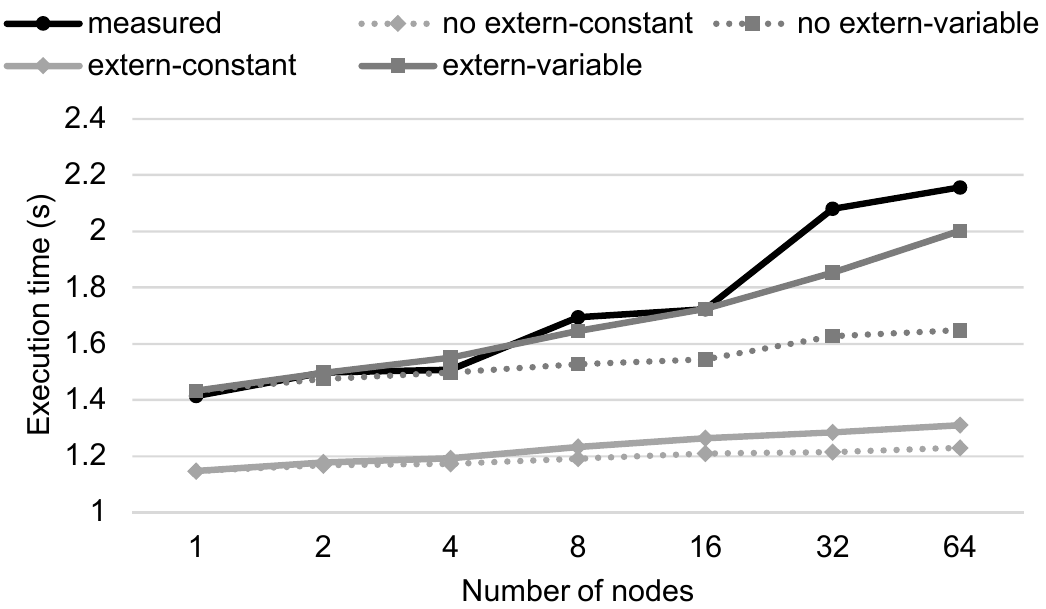}\\
(b) SNAP\\ 
\vspace{4mm}
\includegraphics[width=\columnwidth]{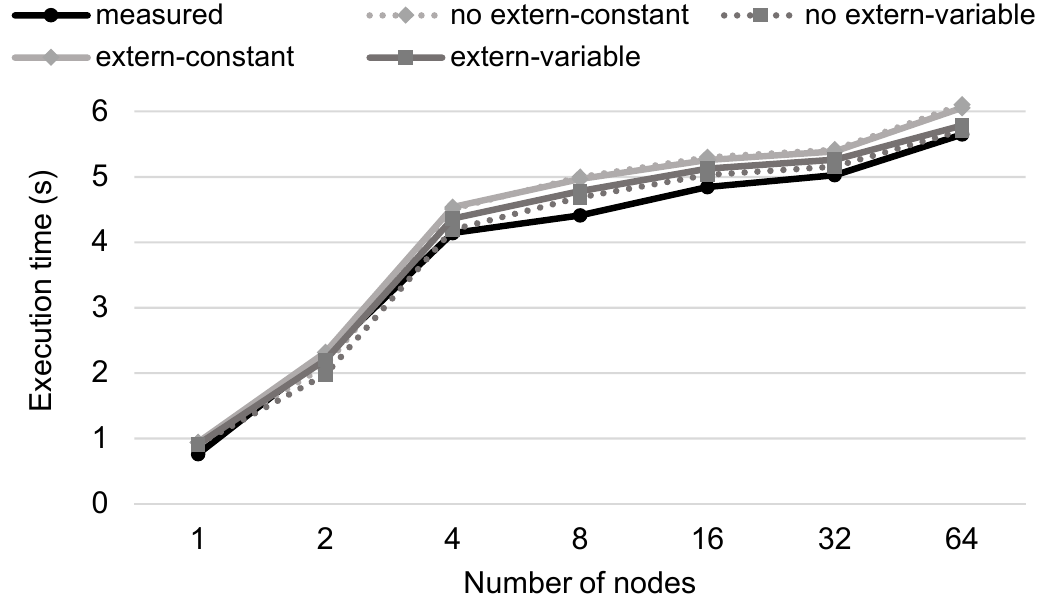}\\
(c) MILC\\
\caption{Measured and estimated execution times when scaling up to 64~nodes. 
}
\label{fig:scaling1}
\end{figure}

\begin{figure}[ptb]
\centering
\includegraphics[width=\columnwidth]{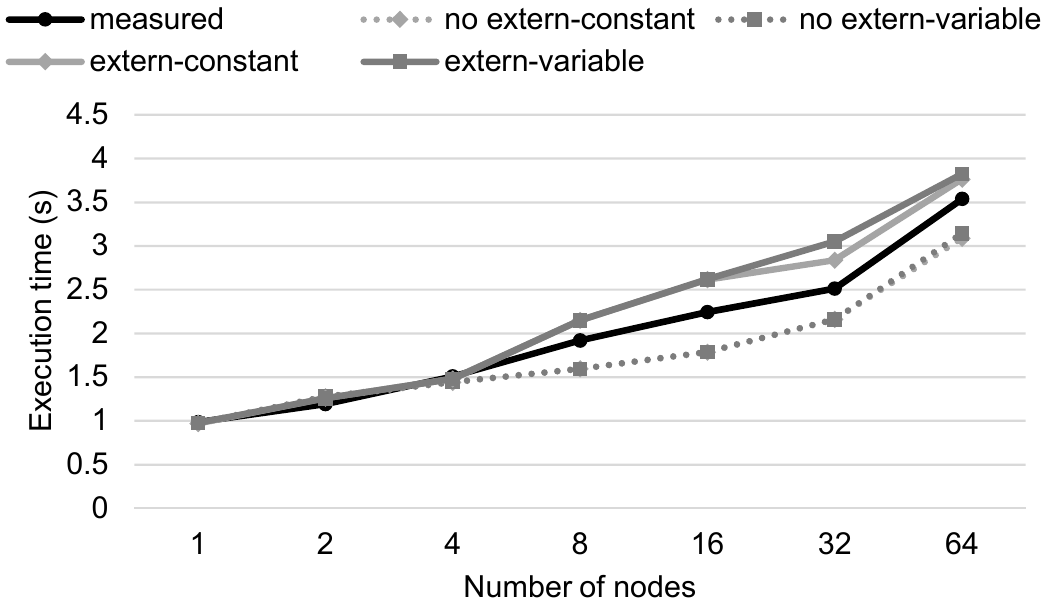}\\
(a) FFT2d \\ 
\vspace{4mm}
\includegraphics[width=\columnwidth]{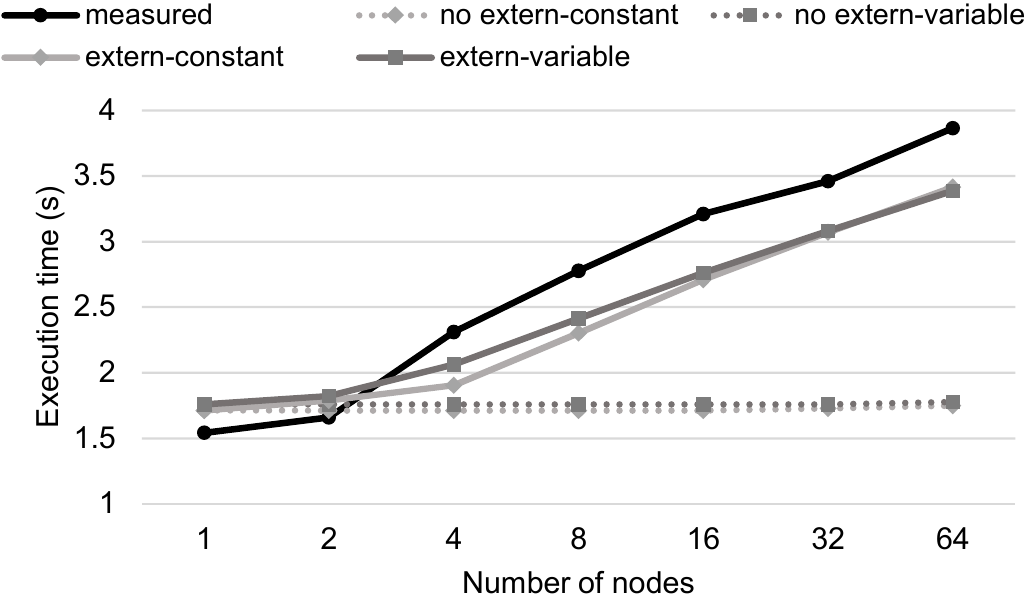}\\
(b) Caffe
\caption{Measured and estimated execution times when scaling up to 64~nodes. 
}
\label{fig:scaling2}
\end{figure}

Figures~\ref{fig:scaling1} and \ref{fig:scaling2} depict the scaling behavior of the five evaluated applications from 1 to 64 nodes.
The curves show the measured execution time, and 4 estimated execution times as described above. All results
are in absolute time (seconds). The black curve corresponds to the measured timings. The dotted curves correspond
to not including external ranks, while the full lines are the results including external ranks. The light gray
curves model a constant timing, and the dark gray curves use variable timings. The dark gray full line curves
correspond to our proposed method.

Many interesting observations can be made from these graphs. We consider single-node results first.
For MILC, FFT2d and Caffe, all projected points at one node are close to the measured timing. This is in fact the error
made by our detailed simulator (5.3\% on average for all applications). 
For HPCG and SNAP, there is a significant difference
between using constant (light gray) and variable (dark gray) timings. In fact, using variable timings reconstructs
the performance reported by our detailed node simulator, while the constant timings underestimate the execution time.
We execute multiple ranks on a node (16 for HPCG and 32 for SNAP), so even for a single node, our multi-node simulator
(SST) simulates the synchronization effects of the MPI operations (assuming a zero communication cost). By assuming
constant timings, we underestimate the synchronization cost, as discussed in Section~\ref{sec:variable}, even for
multiple ranks on a single node. This already shows the importance of using a distribution of timings instead of
constant timings.

Next, we focus on the HPCG results in Figure~\ref{fig:scaling1}(a) as the application scales up to 64~nodes.
Comparing the constant (light gray) and variable
(dark gray) timing curves, it is clear that the variable timings result in more accurate estimates. The difference
between both increases as the number of nodes increases: the more ranks, the larger the difference between the
slowest and the fastest rank. As a result, the variable timings curves follow the trend of the measured results
more closely than the constant timings curves, which are too flat and hence underestimate the scaling
inefficiencies encountered by HPCG.

Adding external ranks (going from the dotted line to the full line) has a noticeable impact when going from
one to two and four nodes. The measured results show an increase in execution time, which is not reflected in the curve
without external ranks (third curve from the top). Adding external ranks faithfully models this rise. Beyond four nodes,
the curve with external ranks runs almost parallel to the curve without external ranks. This can be explained by the
fact that most of the communication of HPCG happens between neighboring ranks. Adding one or three external
nodes adds this local communication, while adding more ranks has little additional impact because there is no communication
between the ranks that are further away. Clearly, the model with our two additions (timing variability and external ranks)
most closely follows the measured results.

We see a similar behavior for SNAP in Figure~\ref{fig:scaling1}(b). The main difference is that the curve with variable
timings and external ranks (second curve from the top) keeps rising at larger node counts (16 to 64 nodes), while the
other curves do not rise that steeply. We notice that adding external ranks not only increases the average compute
time, but also the variability of the compute times, due to extra memory operations that create more cache variability.
More variability results in a longer total execution time, which explains the continuous rise. Although the execution time increase
at 32~and 64~nodes is modeled less accurately, our combined method clearly outperforms the other techniques in terms of
accuracy. 

For MILC, modeling timing variability and external ranks has less impact on the accuracy of the estimations than for
the other benchmarks. MILC is primarily network bandwidth bound, which means that its communication time is mainly determined by network bandwidth, rather than variable compute timings. An
interesting observation is that using constant timings often results in a slightly longer estimated execution time
than using variable timings (the light gray curves are above the dark gray curves). This is just the opposite of what
we see for HPCG and SNAP, where adding variability increases the estimated execution time. The higher execution time
when using constant timings is here caused by the other effect discussed in Section~\ref{sec:variable}: starting
communication at exactly the same time leads to more network congestion and a longer execution time.

For FFT2d we notice that the impact of timing variability is lower than the impact of external ranks. FFT2d
has long compute phases, followed by short intensive communication phases. The variability on the compute part is
rather low, compared to its length. Adding external ranks has a visible impact starting from 8 nodes. The reason is
that although the total number of elements per rank remains constant, the dimensions of the matrix change: for 4 nodes,
each rank performs 16 one-dimensional FFTs on 2,048 elements, while for 8 nodes, each rank calculates 8 1-D FFTs on
4,096 elements. Because FFT has a $n \log(n)$ complexity, execution time increases despite the constant element count.
The impact of adding external ranks is however slightly overestimated, mainly because the computation part is
overestimated by our simulator, see also the next section. Nevertheless, adding external ranks still leads to
the lowest error.

Finally, for Caffe (Figure~\ref{fig:scaling2}), it is also crucial to model the impact of external ranks: without modeling external ranks, the scaling curve is almost flat, which would indicate perfect scaling. However, the measured curve shows an increase in execution time with more nodes, which is tracked more accurately when incorporating the impact of external ranks. When communicating the weights and gradients between the nodes for training the neural network, Caffe encodes the data to reduce its size. At the receiving node, the data needs to be decoded. These encode and decode routines are not called during a single-node execution, so when simulating a single node, these timings are not collected. The encode and decode routines increase the communication latency. Furthermore, the communication is organized as a binary tree, so as more nodes are used, the depth of the communication tree increases. Modeling variable timings has no considerable impact for Caffe, because the computation phases are also rather long compared to their variability. 

Table~\ref{tbl:64nodes} shows the error of the projected execution time at 64 nodes, for the four investigated
methods. The combined method (timing variability and external ranks) is on average the most
accurate method. For some benchmarks, the combined method is not optimal, but the difference with the lowest error falls within the noise of the measurements. 

\begin{table}[!t]
\renewcommand{\arraystretch}{1.1}
\caption{Projection error for each technique at 64 nodes.}
\label{tbl:64nodes}
\centering
\begin{tabular}{lcccc}
\hline
Application & no ext-ct & no ext-var & ext-ct & ext-var\\
\hline
HPCG & 17.0\% & 7.4\% & 15.3\% & 3.5\% \\
SNAP & 43.0\% & 23.6\% & 39.2\% & 7.2\%\\
MILC & 8.0\% & 1.1\% & 7.2\% & 2.4\% \\
FFT2d & 12.7\% & 11.1\% & 6.4\% & 8.1\% \\
Caffe & 54.2\% & 53.5\% & 11.6\% & 12.4\%\\
\hline
Average & 27.0\% & 19.3\% & 15.9\% & 6.7\% \\
\hline
\end{tabular}
\end{table}

\subsection{Time Breakdown Comparison}

We further analyze the accuracy of our technique by breaking down the execution time into computation and
communication components. The compute time is estimated through the detailed simulator measurements,
while the communication time is simulated by the network simulator~(SST). We show that both components
correspond well with measured data, which increases confidence in the proposed technique.

\begin{figure}[tb]
\centering
\begin{tabular}{cc}
\includegraphics[width=0.4\columnwidth]{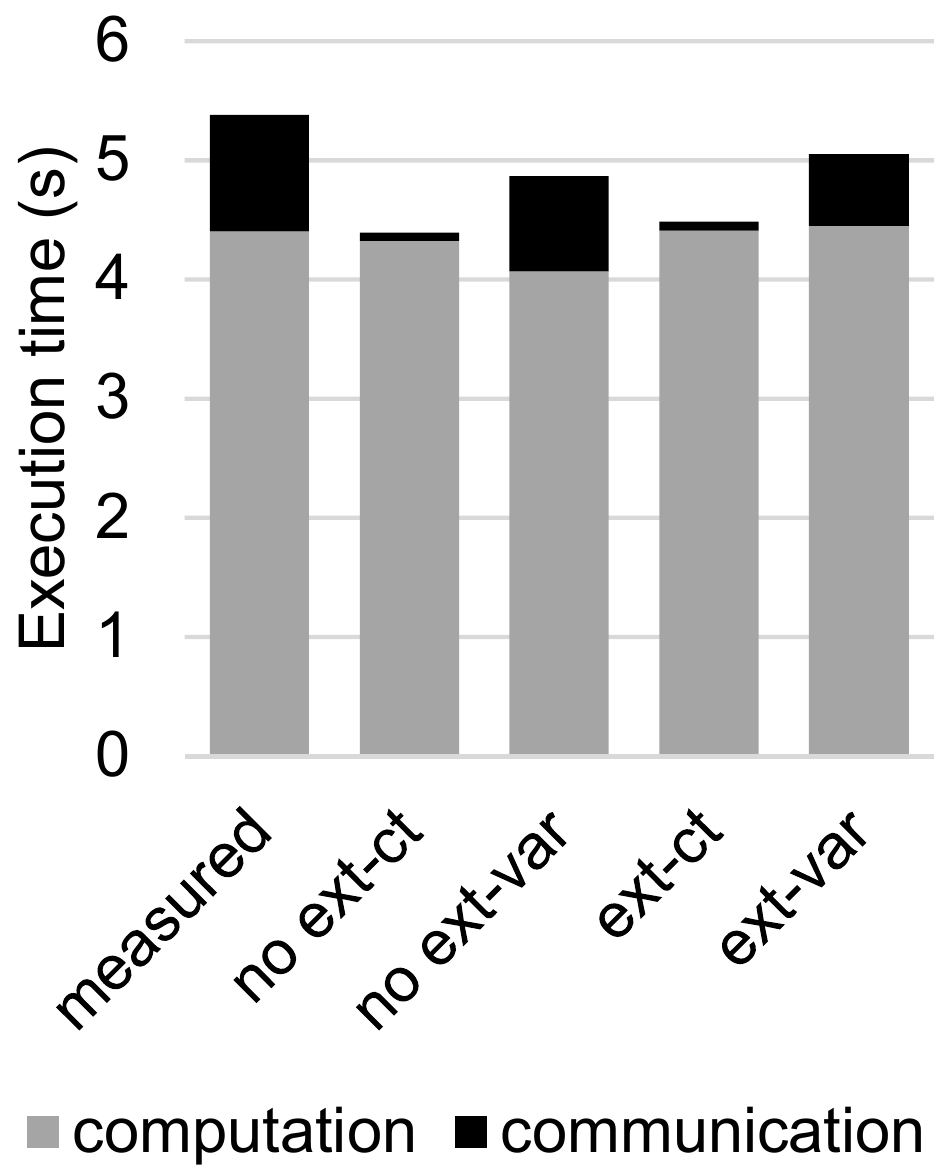} &
\includegraphics[width=0.4\columnwidth]{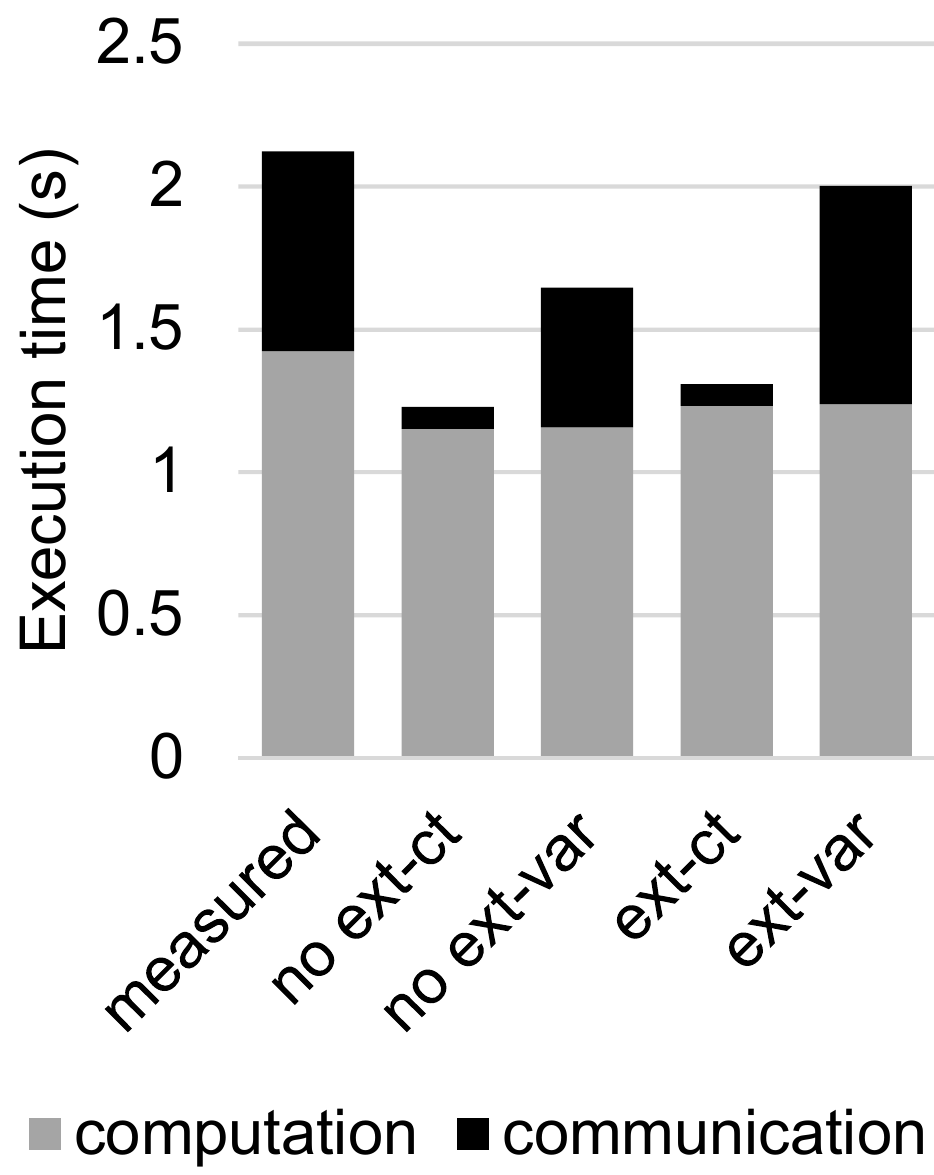}\\
(a) HPCG & (b) SNAP\\
\end{tabular}
\begin{tabular}{ccc}
\includegraphics[width=0.3\columnwidth]{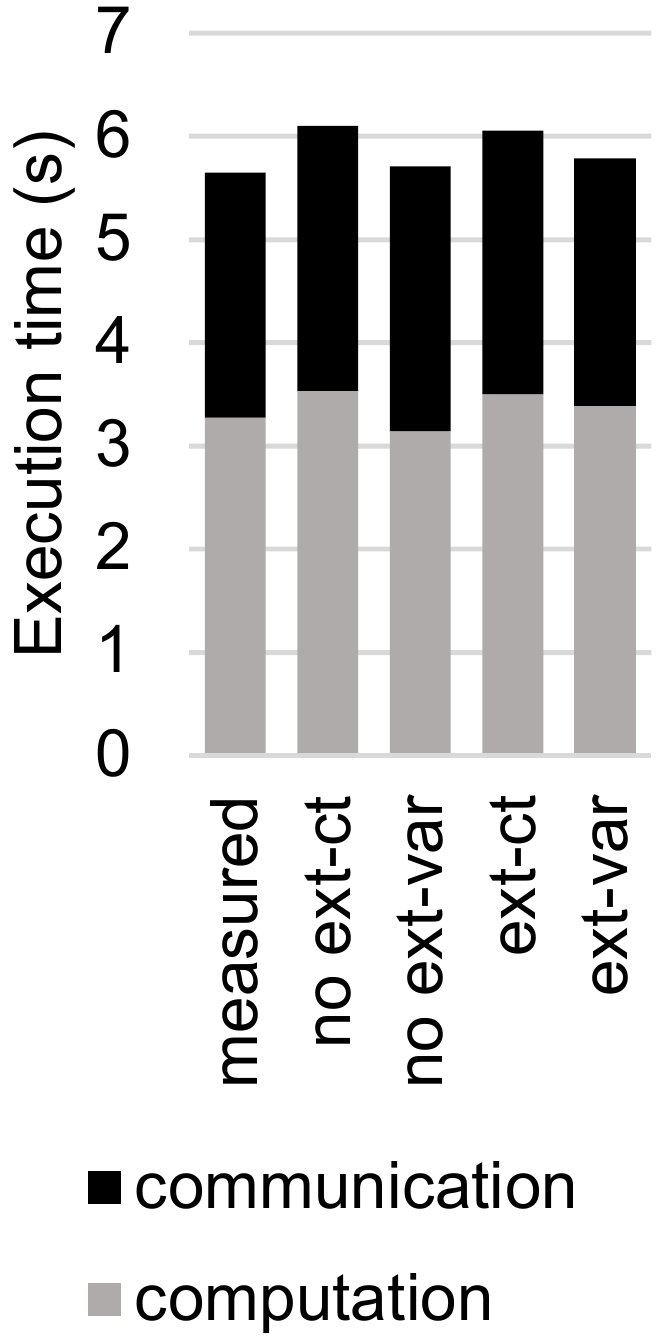} &
\includegraphics[width=0.3\columnwidth]{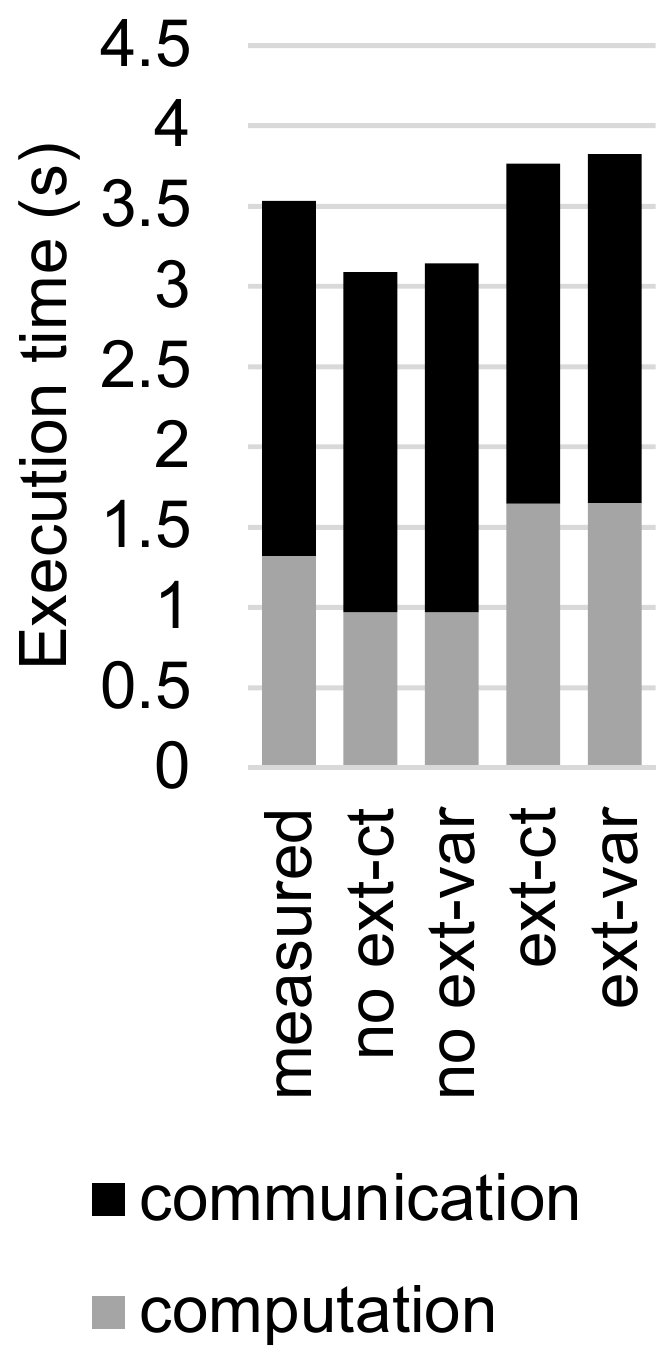} &
\includegraphics[width=0.3\columnwidth]{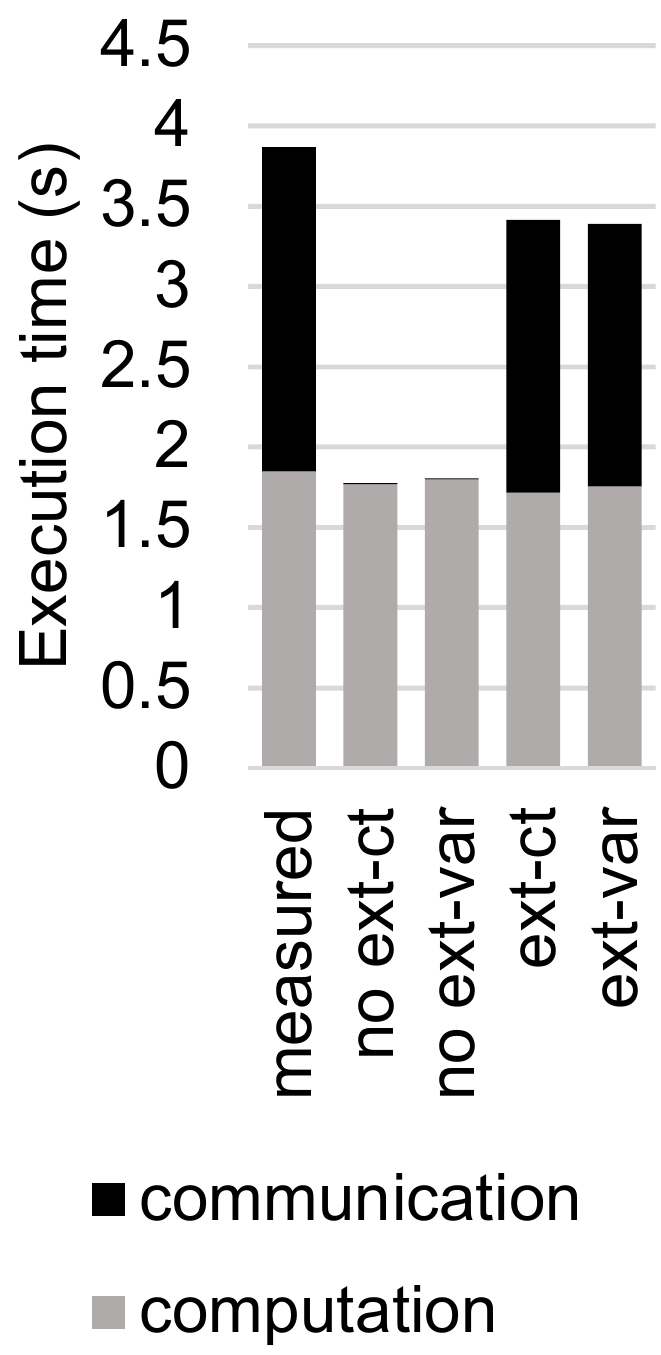}\\
(c) MILC & (d) FFT2d & (e) Caffe
\end{tabular}
\caption{Breakdown into computation and communication times, at 64 nodes.}
\label{fig:components}
\end{figure}

For hardware measurements, we use the Intel MPI Integrated Performance Monitoring (IPM) data collection
tool, which measures the timing of the MPI calls. 
During network simulation (phase 3), we collect the timing of the
MPI calls by inserting timing probes in the event queue of SST. 
Figure~\ref{fig:components} shows the breakdown at
64 nodes for the measured timings and the four evaluated projection techniques. The main observation is that our full
proposal (using timing variability and external ranks; rightmost stack) matches best with the measured timings
(leftmost stack), both in total time (height of the stack) as in the individual components.

For HPCG, it is clear that using constant timings largely underestimates the communication time. The main communication
cost is therefore the variability of the execution time, which makes fast ranks wait on slow ranks. We see an increase
in the computation time when adding external ranks, because of the additional instructions. Our final projection
accurately estimates the computation time, while the communication time is slightly underestimated. This can have
multiple reasons: the actual variability can be even higher due to OS interference (which is not modeled in our
detailed simulator) and/or the actual network bandwidth may be lower than the theoretical bandwidth of 100~Gb/s
due to technical restrictions that are not modeled in our network simulator.

The breakdown for SNAP has a similar behavior. Communication time is increased by modeling variability, and computation
time is slightly increased by adding external ranks. Here, we also see an increase in the communication time after adding
external ranks (compare the third and last graph). Adding external ranks not only increases the number of executed
instructions, but also the variability because of data-dependent behavior. The main error for SNAP is caused by
an underestimation of the computation time, due to modeling abstractions in our detailed simulator.

As expected, the MILC graphs do not show much difference. The communication time for
constant timings is slightly higher than for variable timings, especially with external ranks. As explained before,
this is due to artificial congestion caused by simultaneous network accesses.

For FFT2d, compute time is significantly increased by adding external ranks. Our detailed simulator overestimates
the additional compute time, resulting in a slight overestimation of the total execution time. Communication time is
estimated more accurately, and does not change between projection techniques as the compute time variability is low for this workload.

Caffe shows a considerable increase in modeled communication time when modeling external ranks.
As explained before, this is due to the encode and decode routines that add to the communication latency (we define communication time here as the time during which layers in the neural network are waiting on their new weights).
The main error in the last graph stems from underestimating the communication time.
The compute time variability on the real hardware is larger than in simulation, mainly because it includes reading the training images from disk. Disk read times are not modeled in detail in our simulator, which results in less variable timings. We expect the estimation to be more accurate if a more detailed disk model is used.

\subsection{Microarchitectural Causes of Variability}

\begin{figure}[tb]
\centering
\includegraphics[width=.7\columnwidth]{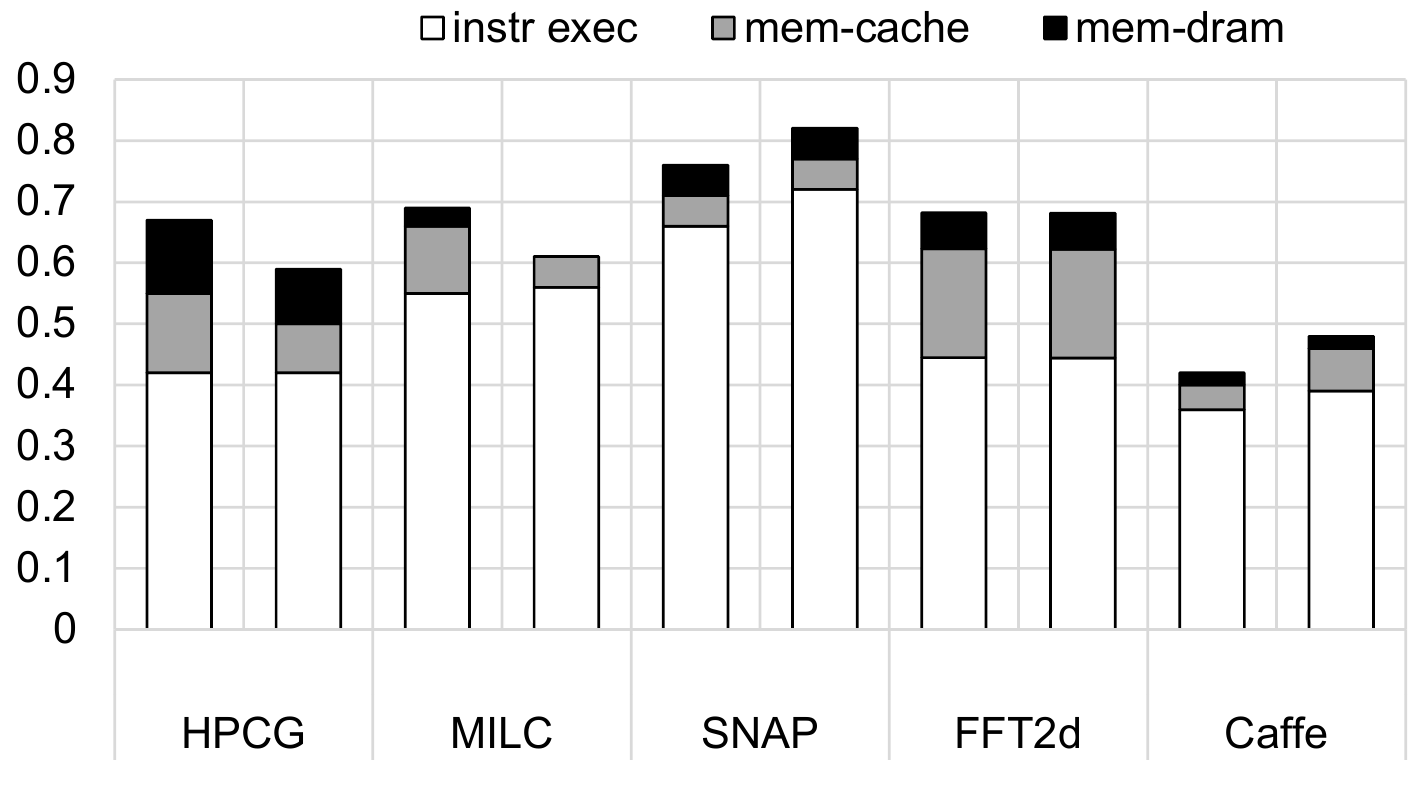}
\caption{CPI stacks of the two most differing ranks or threads all applications.}
\label{fig:stacks}
\end{figure}

To gain more insight into what causes the compute time variability observed in both native execution and detailed simulation,
we consider the simulated CPI (cycles per instruction) stacks for the ranks executed on one node.
A CPI stack divides the CPI of a thread into cycles spent in executing instructions, branch mispredictions, cache and DRAM accesses, etc.
Figure~\ref{fig:stacks} shows the CPI stacks of the two ranks that exhibit
the largest difference in execution time (components that are too small to be visible are not shown).

For HPCG and MILC, we see a similar behavior: the main variability is caused by the cache and DRAM component.
A node consists of two sockets, each containing 18~cores and one shared L3~cache.
Conflicts in the shared L3 cache makes ranks evict each other's data, leading to more misses and DRAM accesses.
Furthermore, the caches are inclusive, meaning that evicted cache lines from L3 are back-invalidated in the L2 and L1 caches,
causing more misses in these levels also.
This behavior depends on the interleaving of cache accesses and the scheduling of the ranks on the sockets, which causes the variability.
Although HPCG and MILC have similar cycle component stacks, the performance of HPCG is much more sensitive to variability than that of MILC, as we saw in the previous sections.
The main reason for this is that MILC is network bandwidth-bound, as we will show in the next section.
As a result, the duration of the communication dominates the effect of timing variability.
Furthermore, the main communication pattern in MILC is point-to-point communication, which means that ranks only have to wait for
their direct communication partners. This is in contrast to bulk-synchronous applications such as HPCG, which use collective communication where all ranks wait on the slowest rank.
Point-to-point communication can therefore overlap much of the compute variability.

The main cause of variability for SNAP is the instruction execution component.
This component is larger than for the other applications.
SNAP makes extensive use of floating point vector operations that are dependent on each other, leading to long instruction latencies and low instruction-level parallelism (ILP).
Furthermore, the inner loop of SNAP contains many data-dependent if-then-else constructs, which leads to divergent code paths.
Because of the low ILP and the long-latency vector instructions, small changes in the code path lead to significant timing differences.

We notice similar behavior for Caffe. Each thread processes a different image of the data set, leading to different instruction counts and different memory behavior (e.g., JPEG decoding, truncation and saturation, etc. lead to data-dependent execution paths).
Finally, as expected, FFT2d has very little variability.

\begin{figure*}[ptb]
\centering
\begin{tabular}{cc}
\includegraphics[width=.7\columnwidth]{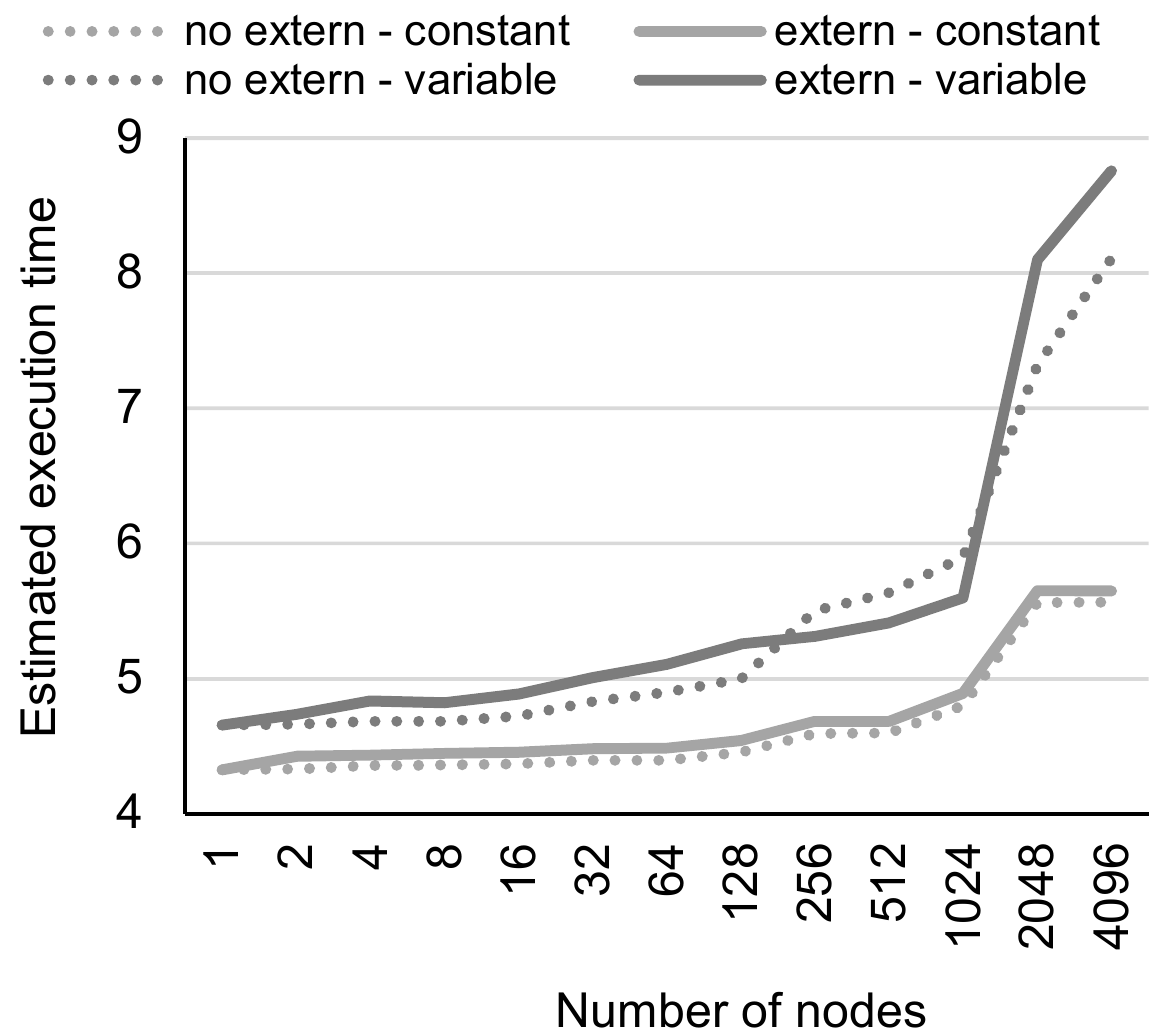} &
\includegraphics[width=.7\columnwidth]{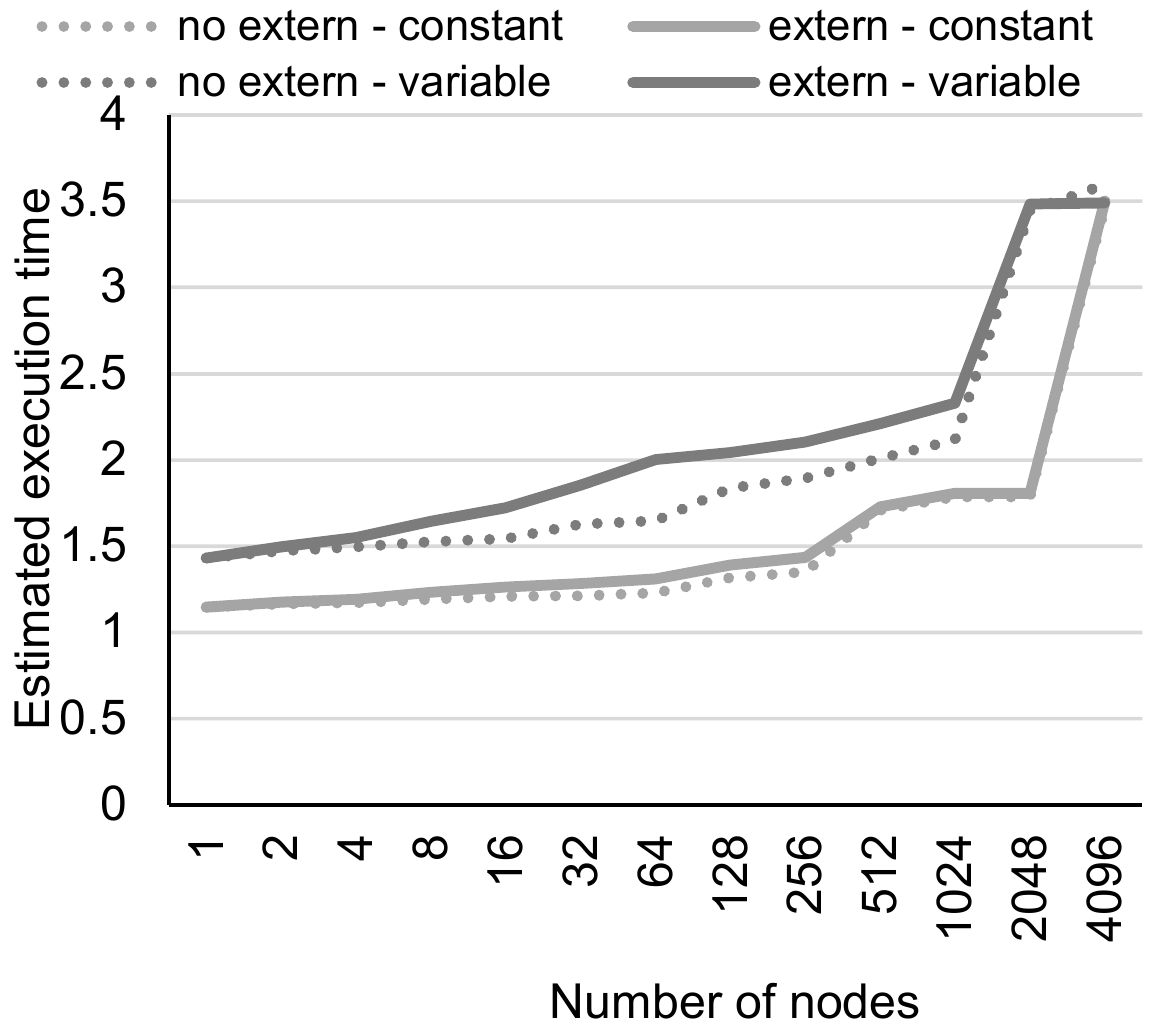} \\
(a) HPCG  & (b) SNAP \\
\includegraphics[width=.7\columnwidth]{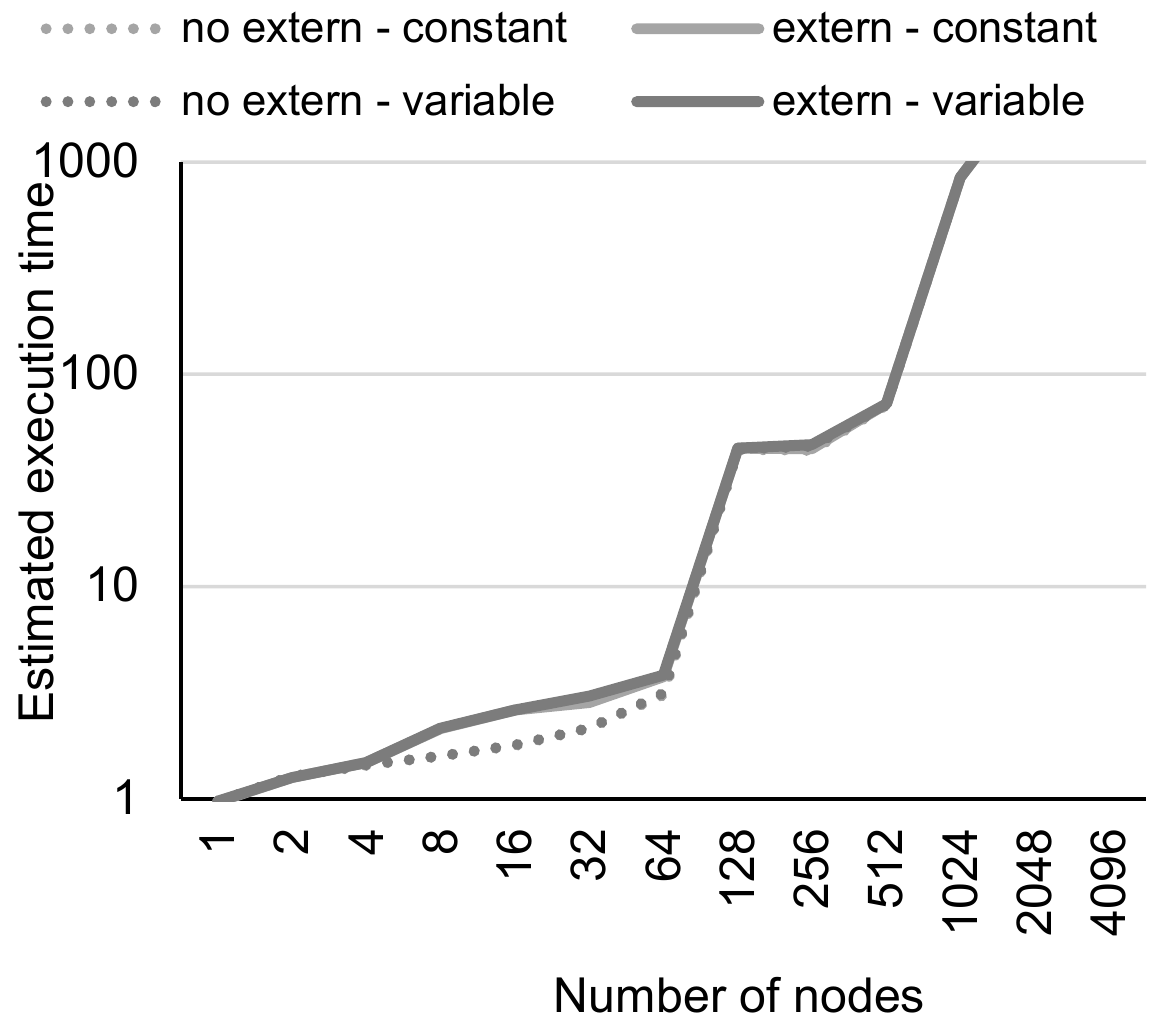} &
\includegraphics[width=.7\columnwidth]{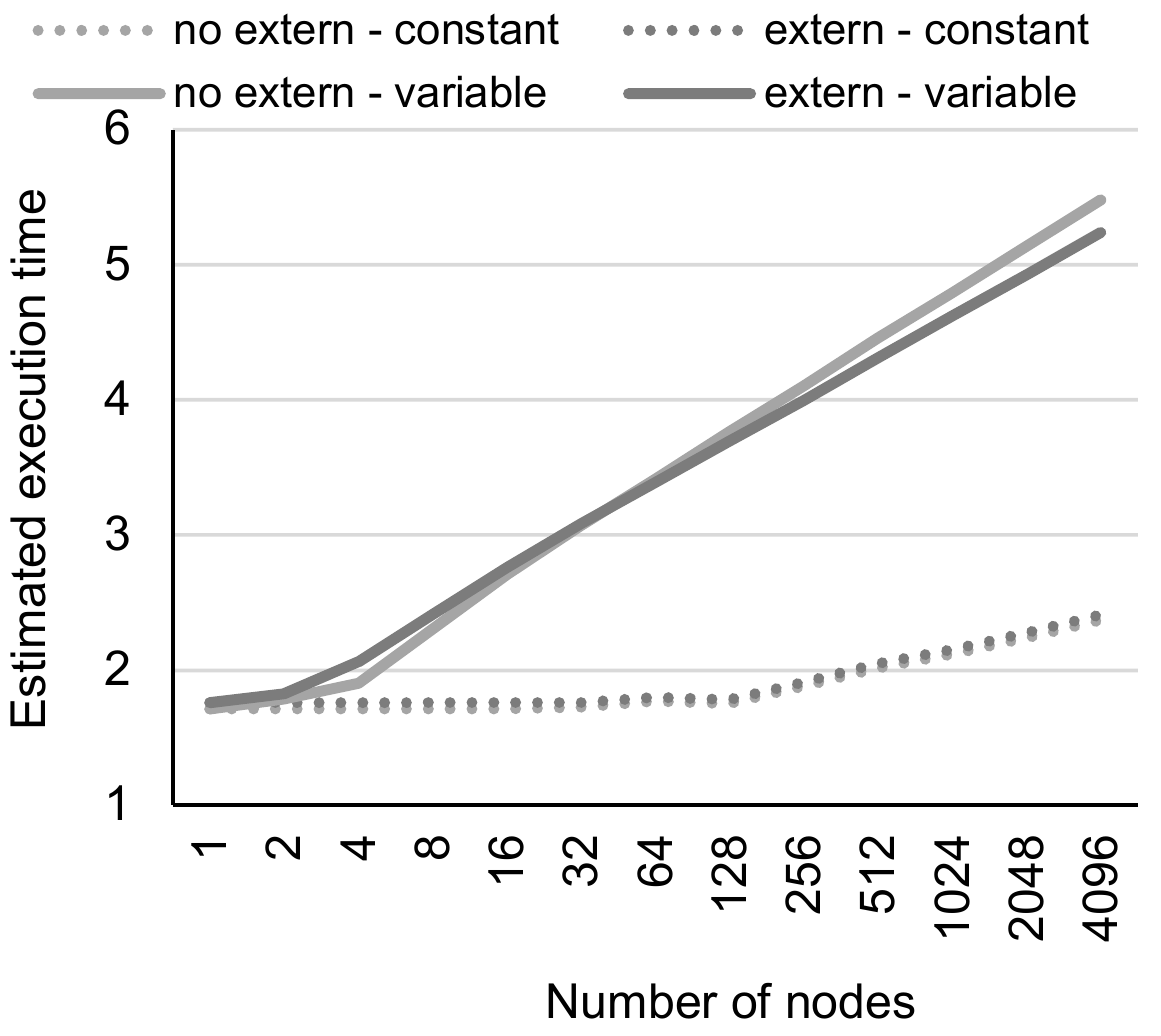} \\
(c) FFT2d (time in log scale) & Caffe\\
\end{tabular}

\caption{Extrapolation of the execution time estimations to 4,096 nodes.}
\label{fig:extrapolation}
\end{figure*}
\section{Case Studies in Design Space Exploration}

Now that we have validated our technique by comparing its predictions to hardware measured timings, we
show some examples of studies that can be performed using our method. We extrapolate our predictions
to 4,096 nodes, and show how our results can be used to scale the network configuration.

\subsection{Extrapolation to Many Nodes}

Once we have a communication skeleton and the timings of the computation parts, we can easily extrapolate
the predicted execution time to many nodes using SST simulations.
Figure~\ref{fig:extrapolation} shows the predicted execution
times for three of our five applications up to 4,096 nodes. 
The other two applications, MILC and FFT2d are already bandwidth limited at 64 nodes, and because their communication scales superlinearly with the number of nodes, their execution time at large node counts consists mainly of network transfer time, which is the same for all techniques. 
Therefore, we leave them out to save space.

We again show
the four evaluated techniques, but now without a hardware measured curve, because we do not have access to
that many nodes. The main conclusion from these results is that the differences between the predicted
execution times using the different techniques increase even more as we increase the number of nodes.
For example, for HPCG, the difference between using constant timings without external ranks, and using
variable timings with external ranks is almost 60\%.

Another interesting observation is that using constant timings does not always capture behavior that our
most accurate method (\emph{external -- variable}) predicts. For example, for HPCG and SNAP, we see a big
jump in the predicted execution time when increasing the number of nodes from 1,024 to 2,048 for \emph{external -- variable}.
Using constant timings predicts a much smaller increase for HPCG, and almost no increase for SNAP.
At 2,048 nodes, the bandwidth requirements for both HPCG and SNAP come close to the available bandwidth
(communication traffic scales superlinearly with the number of nodes). As a result, the network gets congested,
leading to longer communication times. Variable timings reinforce this effect: quicker ranks access the network
early, and see almost no congestion, while slower ranks are slowed down even more, because they have to compete
with the messages of the faster ranks. For SNAP, the estimations converge at 4,096 nodes, because the bandwidth
requirements exceed the provisioned bandwidth by a significant amount. The execution time is now largely determined by the
bandwidth bottleneck, `absorbing' the timing variability effects. Nevertheless, using constant timings
predicts this to happen between 2,048 and 4,096 nodes, while our most accurate method indicates that bandwidth
saturates earlier, between 1,024 and 2,048 nodes.

The communication time underestimation of not modeling external ranks for Caffe continues to increase as the node count increases. The difference between extern-variable and no extern-constant increases to 55\%. The logarithmic scaling behavior (note the logarithmic X-axis) is caused by the tree-based reduction: the longest path in the tree is logarithmic in the number of nodes.

\subsection{Network Configuration Sensitivity}

\begin{figure}[tb]
\centering
\includegraphics[width=.9\columnwidth]{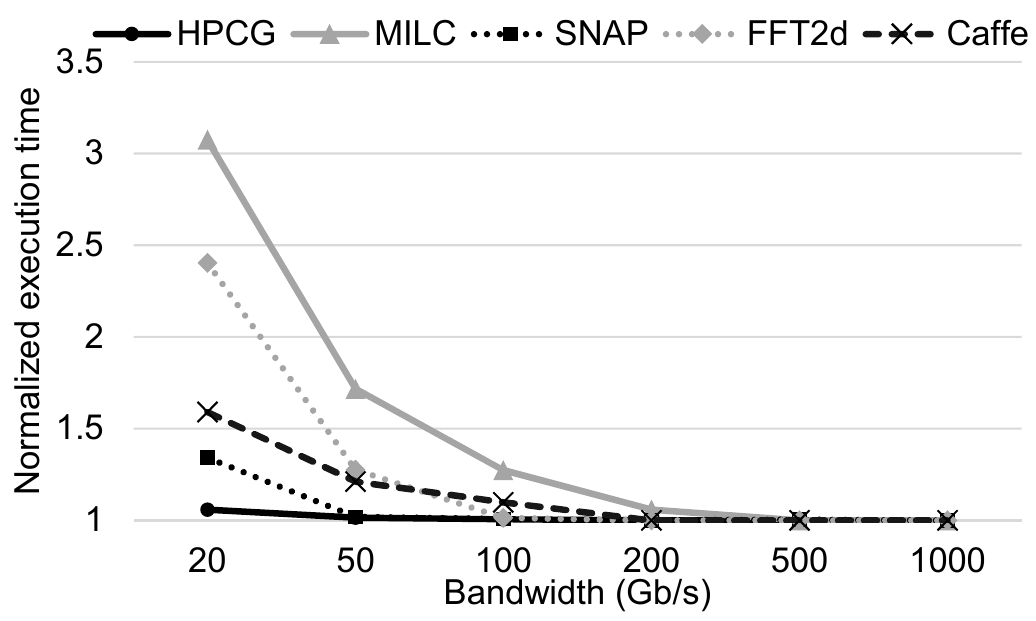}
\caption{Estimated execution time on 64 nodes for varying network bandwidth (normalized to 1000~Gb/s).}
\label{fig:bandwidth}
\end{figure}

\begin{figure}[tb]
\centering
\begin{tabular}{cc}
\includegraphics[width=0.45\columnwidth]{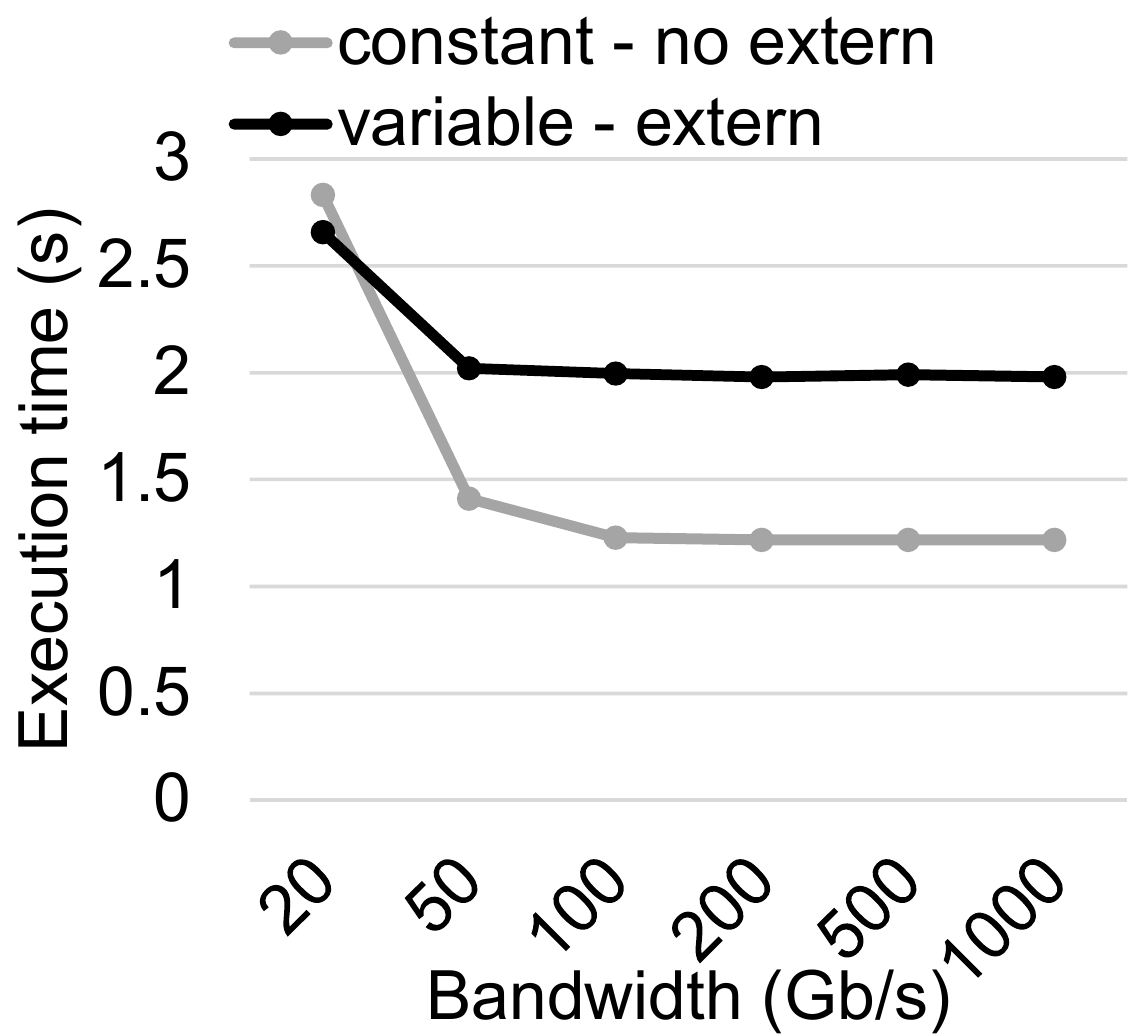} &
\includegraphics[width=0.45\columnwidth]{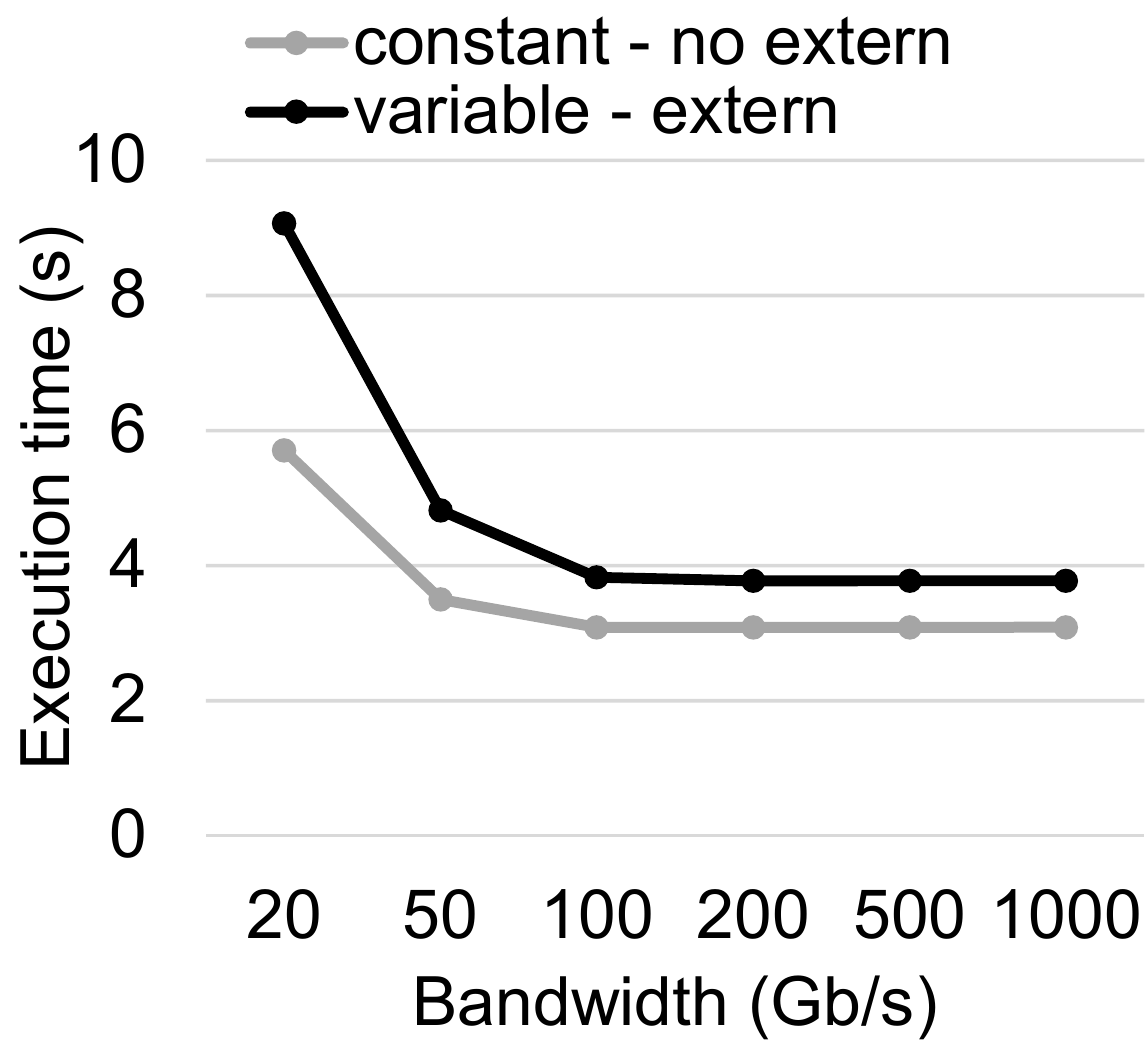} \\
(a) SNAP & (b) FFT2d
\end{tabular}
\caption{Network bandwidth behavior for SNAP and FFT2d on 64 nodes, using constant timings without external ranks, and variable timings with external ranks.}
\label{fig:bandwidth_SNAP_FFTW}
\end{figure}

We now vary network bandwidth and network topology, projecting total execution time for different network
configurations. We configure SST to simulate network bandwidths of 20, 50, 100 (default), 200 500, and 1000~Gb/s, and network link latencies (between node and router, and between routers) of 40, 90 (default), 140, 200, 500 and
1000~ns.  
No new detailed node simulations are needed.

Figure~\ref{fig:bandwidth} shows the execution time on 64 nodes as a function of network bandwidth, using our most accurate technique (ext-var), with the default 90~ns link latency. 
Times are normalized to 1000~Gb/s. This figure illustrates
how our technique can be used to dimension the network: for HPCG, a 20~Gb/s network is sufficient, while
MILC still profits from going to 200~Gb/s and even 500~Gb/s (although marginally). The other applications
are in between. 
Note that for Caffe, the slope of decreasing execution time as a function of bandwidth is smaller than for the other applications.
This is because the communication time consists of two parts: the data transfer time, which is determined by bandwidth, and the encode-decode time, which is independent of the bandwidth.
Therefore, bandwidth increases for Caffe have a smaller impact on communication time reduction than for other applications.

The importance of using the correct timing model is illustrated in Figure~\ref{fig:bandwidth_SNAP_FFTW}.
If we use constant timings without external ranks for SNAP, we find that going from 20~Gb/s to 50~Gb/s
halves the execution time, while in reality, the reduction is only 24\%. Constant timings overestimate the
execution time at 20~Gb/s by modeling too much congestion caused by concurrent network accesses, while it underestimates the
execution time at 50~Gb/s because the extra synchronization due to variable timings is not captured.
Using constant timings even suggests a small performance improvement at 100~Gb/s, which is not present
in our most accurate model: the small send time reduction is completely absorbed by the synchronization
penalty.

The opposite behavior is seen for FFT2d between 20~Gb/s and 50~Gb/s: the model using constant timings
predicts an execution time reduction of 2.2~s, while our most accurate model predicts a 4.3~s reduction
of execution time, almost doubling the expected performance improvement. In this case, the bandwidth saturation at 20~Gb/s reinforces
the variability and thus the synchronization time, as discussed earlier. This extra
synchronization time is not modeled by using constant timings. Note that Figure~\ref{fig:scaling1}(d)
showed little impact of adding variable timings at 100~Gb/s for FFT2d, but when the bandwidth is more
restricted, variability plays a much more significant role.

\begin{figure}[t]
\centering
\includegraphics[width=.9\columnwidth]{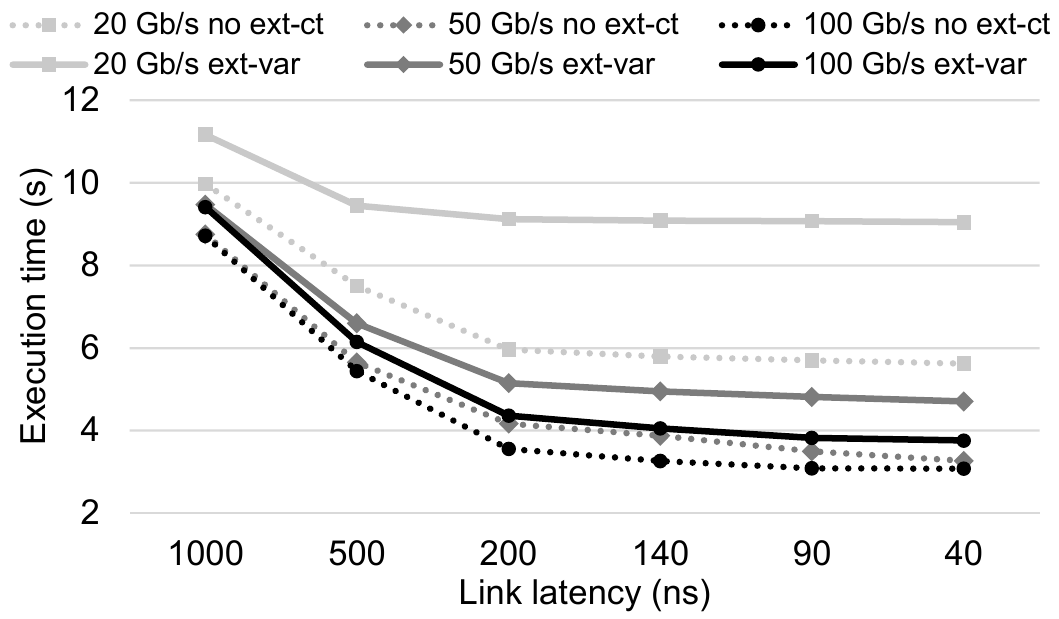}
\caption{Estimated time for FFT2d on 64 nodes, for different network bandwidths and link latencies.}
\label{fig:latency_FFTW}
\end{figure}

Figure~\ref{fig:latency_FFTW} shows the dependence of FFT2d on link latency (the other applications have no
significant link latency dependent execution time). Clearly, we cannot improve execution time much beyond
the default 90~ns link latency. The figure is more interesting at higher link latencies, and shows the
correlation between bandwidth and latency. Our most accurate estimations (\emph{ext-var}) shows that at low bandwidths (e.g., 20~Gb/s) the application is less
sensitive to the link latency (minimum execution time is already obtained at 500~ns latency). For
high bandwidths, link latency is a more determining factor: at 100~Gb/s, we need a link latency of 90~ns
to become within 5\% of the optimal execution time. 

Once more, using constant timings and no external ranks leads to wrong conclusions: 
at 20 Gb/s, it predicts an execution time reduction of 21\% when link latency reduces from 500~ns to 200~ns, while in fact it is only 3.5\%. 
A similar observation can be made at 50 Gb/s: using constant timings predicts performance improvements up to 40~ns, while performance saturates at 140~ns.
In these cases, the latency reduction is overlapped by the synchronization due to variable timings, which is not modeled by using constant timings.

Lastly, we experiment with different network topologies. 
Our default topology is a fat tree, with 24 nodes at each leaf switch (each switch has 48 ports, i.e., the number of ports on a high-end Intel OmniPath switch).
To support up to 2,048 nodes, we instantiate 96 leaf switches (4 times 24), 96 second level switches, and 48 root switches, for a total of 240 switches.
Other topologies are more efficient in terms of hardware cost, at the penalty of not having exclusive channels between each pair of nodes.
We evaluate a 3D torus with 100 switches (5x5x4) and a dragonfly network~\cite{dragonfly}, with 90 switches (5 groups of 18 switches).
Both configurations have 24 compute nodes per switch, leaving 24 ports for communication.
For the torus network, we use 4 links for each of the 6 directions, and for the dragonfly, 17 links are used to connect to switches in the same group, and 7 for inter-group connections.

\begin{figure}[tb]
\centering
\includegraphics[width=.9\columnwidth]{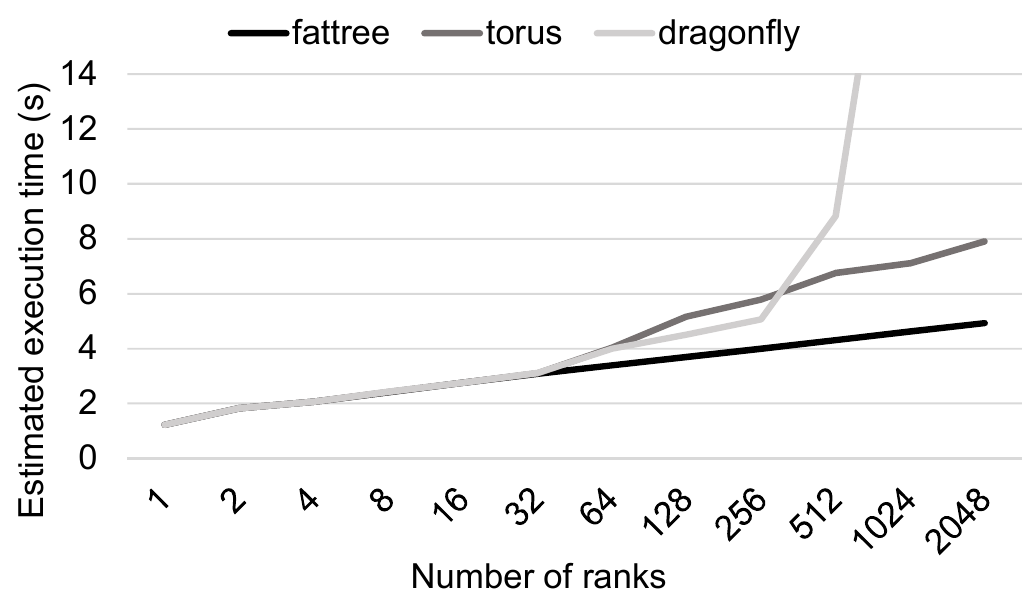}
\caption{Estimated execution time of Caffe for 1 to 2,048 nodes with three different network topologies.}
\label{fig:topology}
\end{figure}

Figure~\ref{fig:topology} shows the projected execution time (using our most accurate technique) for Caffe on the three topologies from 1 to 2,048 ranks (1 rank per node).
Up to 32 ranks, there is no difference between the topologies: in each configuration, 24 nodes are attached to the same switch, and the 8 ranks that go on another switch at 32 ranks do not generate enough traffic to stress the difference in bandwidth between the switches.
Beyond 32 nodes, the fat tree clearly outperforms the other two configurations (at the cost of almost 3$\times$ the number of switches).
Comparing the torus and dragonfly topology, we see that initially, dragonfly performs better than torus because it has more connections between switches in the same group.
As soon as multiple groups are needed (from 512 nodes, which is larger than $18 \times 24$), torus performs better, meaning that the inter-group connections become a bottleneck for dragonfly. 

\section{Conclusion}

We propose a simulation methodology for many-node architectures that targets both accuracy and scalability,
which are often conflicting requirements. Our method combines the profiling speed of native execution, the accuracy
of detailed processor simulation, and the scalability of high-level network simulation. We find that
microarchitecture-dependent timing variability has a significant impact on many-node performance, as well
as instruction and memory operation overhead due to problem size scaling. By including timing variability
models and by adding natively running external ranks to a single-node simulation, we are able to predict the
performance of a many-node application to within 6.7\% on average (12\%~maximum error), compared to an average 27\%~error and up to
54\% without these additions, which is the state-of-the art method. At larger node counts, the difference between the timing projections through
modeling and not modeling these additions, continues to increase.
Furthermore, not including timing variability and external communication overhead in design space exploration results in selecting the wrong optimal configuration, and potentially procuring the wrong machine.

By separating single-node detailed simulation and high-level network simulation, we are able to scale to
thousands of nodes, while requiring only a minimum of compute and memory resources.  The technique can be
used to dimension detailed parameters of the nodes (microarchitectural configuration) and network
(topology, latency and bandwidth).

\section*{Acknowledgements}
%We thank all reviewers for their insightful comments and suggestions.
We thank Pardo Keppel for proofreading this paper and for his useful suggestions to improve it.

\bibliographystyle{IEEEtranS}
\bibliography{paper}

% that's all folks
\end{document}